--------------------------- cut here ----------------------------------------

\documentstyle[12pt]{article}
\catcode`@=11
\def\seceqa{\@addtoreset{equation}{section}
           \def\theequation{1.\arabic{equation}}}
\def\seceqb{\@addtoreset{equation}{section}
           \def\theequation{2.\arabic{equation}}}
\def\seceqc{\@addtoreset{equation}{section}
           \def\theequation{3.\arabic{equation}}}
\def\seceqd{\@addtoreset{equation}{section}
           \def\theequation{4.\arabic{equation}}}
\def\seceqe{\@addtoreset{equation}{section}
           \def\theequation{5.\arabic{equation}}}
\def\seceqf{\@addtoreset{equation}{section}
           \def\theequation{6.\arabic{equation}}}
\def\seceqg{\@addtoreset{equation}{section}
           \def\theequation{7.\arabic{equation}}}
\def\seceqaa{\@addtoreset{equation}{section}
           \def\theequation{A\arabic{equation}}}
\def\seceqab{\@addtoreset{equation}{section}
           \def\theequation{B\arabic{equation}}}
\catcode`@=12

\begin{document}

\title{Relativistic Coulomb Sum Rules for $(e,e^\prime)$}
\author{T.C.~Ferr\'ee and D.S.~Koltun \\
{\it Department of Physics and Astronomy}\\
{\it University of Rochester, Rochester, NY 14627-0171}}
\maketitle
\vskip 0.5 true in

\begin{abstract}
A Coulomb sum rule is derived for the response of nuclei to $(e,e^\prime)$
scattering with large three-momentum transfers.  Unlike the nonrelativistic
formulation, the relativistic Coulomb sum is restricted to spacelike
four-momenta for the most direct connection with experiments; an immediate
consequence is that excitations involving antinucleons, e.g., $N{\bar N}$
pair production, are approximately eliminated from the sum rule.
Relativistic recoil and Fermi motion of target nucleons are correctly
incorporated.   The sum rule decomposes into one- and two-body parts,
with correlation information in the second.  The one-body part requires
information on the nucleon momentum distribution function, which is
incorporated by a moment expansion method.  The sum rule given through
the second moment (RCSR-II) is tested in the Fermi gas model, and is
shown to be sufficiently accurate for applications to data.
\end{abstract}

\noindent PACS number(s): 25.30.-c, 25.30.Fj, 24.10.-i, 11.50.Li
\vskip 0.5 true in
\centerline{Submitted to {\bf Physical Review C}}
\vfill\eject

\setcounter{equation}{0}
\seceqa
\section{Introduction}

Electrons have proven to be useful probes for the study of nuclear structure,
primarily because the $eN$ interaction is reasonably well-known and because
the electromagnetic interaction is weak compared to the strong forces which
dominate nuclear structure.  The longitudinal contribution to the nuclear
response is of particular interest because of its relationship to two-body
correlation functions through what is commonly known as the Coulomb sum rule
(CSR).\cite{mvvh}  With the assumption that the nucleus may be treated
as a nonrelativistic system of interacting nucleons, the extraction of
nucleon-nucleon correlations is formally direct: this is commonly referred
to as the nonrelativistic Coulomb sum rule (NRCSR).  In practice, however, one
finds that to obtain useful information one must extend the experiments and
analysis to energies and momentum transfers sufficiently large that the usual
nonrelativistic assumptions fail.  This will be particularly the case for
experiments at the Continuous Electron Beam Accelerator Facility (CEBAF)
with $E_{\rm beam}\!\simeq\!{4\rm\ GeV}$.  Therefore, it has become desirable
to have a fully-relativistic formalism both for analyzing experimental data
and for testing theoretical models of nuclear structure.  In this paper, we
present an extension of the NRCSR which successfully handles certain problems
which arise in electron scattering from nuclei at large three-momentum
transfers.

Before discussing further the goals of this paper, it will be useful to
review the form of the NRCSR, as well as its derivation.  We begin with
the first-order Born (or one-photon exchange) approximation for the
scattering of ultrarelativistic electrons ($|{\bf k}|\!>\!\!>m_e$) from
nuclear targets, in which one can factor the $(e,e^\prime)$ differential
cross-section into leptonic and nuclear parts.  After performing a standard
separation of this nuclear response function into longitudinal and
transverse (virtual) photon contributions, the differential cross-section
in the laboratory frame can be written

	\begin{equation}{d^2\sigma\over d\Omega^\prime dE^\prime}=
	{d\sigma\over d\Omega^\prime}\biggr\vert_{\rm Mott}
	\biggl[{Q^4\over{\bf q}^4}W_C(\omega,{\bf q})+\biggl({1\over2}
	{Q^2\over{\bf q}^2}+{\rm tan}^2{\theta\over2}\Bigr)
	W_T(\omega,{\bf q})\biggr],\label{aa}\end{equation}

\noindent where the Mott cross-section

	\begin{equation}{d\sigma\over d\Omega^\prime}
	\biggr\vert_{\rm Mott}\equiv{\alpha^2{\rm cos}^2(\theta/2)
	\over4E^2{\rm sin}^4(\theta/2)}\label{ab}\end{equation}

\noindent
describes the elastic scattering of ultrarelativistic electrons from a
{\it fixed\/} point target with charge $e$ and no spin.  Here $q^\mu\!=
(\omega,{\bf q})$ is the four-momentum transfer to the target, $\theta$
is the electron scattering angle, $E$ is the electron beam energy and
$Q^2\!\equiv\!-q^2\!=\!{\bf q}^2\!-\!\omega^2$.  The first term in the
square brackets in (\ref{aa}) is the longitudinal contribution, and is
usually expressed in terms of the Coulomb response function:

	\begin{equation}W_C(\omega,{\bf q})\equiv\sum_f
	|\langle f|{\hat\rho}(q)|i\rangle|^2
	\ \delta(\omega-E_f+E_i),\label{ac}\end{equation}

\noindent where $|i\rangle$ and $|f\rangle$ denote initial\footnote
{For notational simplicity, we assume a nondegenerate ground state
$|i\rangle$.  The results can easily be generalized to unpolarized
targets with $J\!\not=\!0$.} and final nuclear states, respectively,
and ${\hat\rho}(q)$ is the Fourier transform of the nuclear charge
density operator.

In general, the operator ${\hat\rho}(q)$ should include contributions from
both the nucleons and (virtual) charged mesons in the target, but it has
been conventional in nonrelativistic approximations to ignore the latter.
Then ${\hat\rho}(q)$ is given by the spatial Fourier transform of the
local nucleon charge density operator:

	\begin{equation}\hat\rho(q)\equiv G_{E,p}(Q^2)
	\int d^3x\ {\rm e}^{i{\bf q\cdot x}}\
	{\hat\psi}^\dagger({\bf x}){\hat Q}
	{\hat\psi}({\bf x}),\label{ad}\end{equation}

\noindent where ${\hat\psi}({\bf x})$ is the local (nonrelativistic) nucleon
field-operator, $\hat Q\!=\!(1\!+\!\tau_3)/2$ is the charge operator in
isospin space which projects out protons, and $G_{E,p}(q^2)$ is the proton
charge form factor.  (In a purely nonrelativistic treatment of the target,
$G_{E,p}(Q^2)$ would be replaced by the Fourier transform of the proton
charge density $F_p({\bf q}^2)$.)  Charge effects of neutrons are small
and are usually neglected.

To proceed to a sum rule, we first define the nonrelativistic Coulomb sum
function for inelastic scattering at fixed three-momentum transfer {\bf q}:

	\begin{eqnarray}
	S_{NR}({\bf q}) & \equiv & \int
	_{\omega_{el}^+}^\infty d\omega\ {W_C(\omega,{\bf q})
	\over G_{E,p}^2(Q^2)}\nonumber\\
	&  & \nonumber\\
	& = & \int d^3x d^3x^\prime\ {\rm e}^{i{\bf q}\cdot
	({\bf x}-{\bf x}^\prime)}\sum_{f\not=i}
	\langle i|{\hat\psi}^\dagger({\bf x}^\prime){\hat Q}
	^\dagger{\hat\psi}({\bf x}^\prime)|f\rangle\langle f|
	{\hat\psi}^\dagger({\bf x}){\hat Q}{\hat\psi}({\bf x})
	|i\rangle,\label{ae}\end{eqnarray}

\noindent where the integration is over all energies above the elastic
peak at $\omega\!=\!\omega_{el}$.  Expression (\ref{ae}) then becomes a
sum rule by subtracting the elastic term $f\!=\!i$ explicitly and using
the closure relation

	\begin{equation}\sum_f|f\rangle\langle f|=1,\label{af}
	\end{equation}

\noindent on the resulting sum over nuclear states $|f\rangle$. Equation
(\ref{ae}) is then in the form of a Fourier transform of the expectation value
of four nucleon field operators.  To separate out the two-body correlation
part, we put the operators in normal order (creation operators to the left)
by anticommutation, using

	\begin{equation}\lbrace{\hat\psi}({\bf x}),{\hat\psi}^\dagger
	({\bf x}^\prime)\rbrace=\delta({\bf x}\!-\!{\bf x}^\prime).
	\label{ag}\end{equation}

\noindent The sum rule then becomes

	\begin{eqnarray}S_{NR}({\bf q}) & = & \int d^3x\ \langle i|
	{\hat\psi}^\dagger({\bf x}){\hat Q}{\hat\psi}({\bf x})
	|i\rangle\nonumber\\ &   & \nonumber\\
	& + &\int d^3x d^3x^\prime\
	{\rm e}^{i{\bf q}\cdot({\bf x}-{\bf x}^\prime)}
	\biggl\lbrace\langle i|{\hat\psi}^\dagger({\bf x})
	{\hat Q}{\hat\psi}^\dagger({\bf x}^\prime)
	{\hat\psi}({\bf x}^\prime){\hat Q}{\hat\psi}({\bf x})
	|i\rangle\nonumber\\
	&  & \nonumber\\
	&  & \qquad\qquad\qquad
	-\langle i|{\hat\psi}^\dagger({\bf x}){\hat Q}{\hat\psi}
	({\bf x})|i\rangle\langle i|{\hat\psi}^\dagger
	({\bf x}^\prime){\hat Q}{\hat\psi}({\bf x}^\prime)
	|i\rangle\biggr\rbrace.\label{ah}\end{eqnarray}

\noindent The first term is simply the total nuclear charge $Z$, and
the second term can be written in terms of a correlation function:

	\begin{eqnarray}C_{NR}({\bf q})
	& = & \int d^3x d^3x^\prime\
	{\rm e}^{i{\bf q}\cdot({\bf x}-{\bf x}^\prime)}
	\biggl\lbrace\langle i|{\hat\psi}^\dagger({\bf x})
	{\hat Q}{\hat\psi}^\dagger({\bf x}^\prime)
	{\hat\psi}({\bf x}^\prime){\hat Q}{\hat\psi}({\bf x})
	|i\rangle\nonumber\\
	&  & \nonumber\\
	&  & \qquad\quad
	-\biggl({Z-1\over Z}\biggr)\langle i|
	{\hat\psi}^\dagger({\bf x}){\hat Q}{\hat\psi}({\bf x})
	|i\rangle\langle i|{\hat\psi}^\dagger({\bf x}^\prime)
	{\hat Q}{\hat\psi}({\bf x}^\prime)
	|i\rangle\biggr\rbrace,\label{ai}\end{eqnarray}

\noindent and the elastic form factor of the nuclear target:

	\begin{equation}F_{el}({\bf q})=G_{E,p}(Q_{el}^2)
	\int d^3x\ {\rm e}^{i{\bf q}\cdot{\bf x}}\ \langle i|
	{\hat\psi}^\dagger({\bf x}){\hat Q}{\hat\psi}({\bf x})
	|i\rangle,\label{aj}\end{equation}

\noindent where $Q_{el}^2\!\equiv\!{\bf q}^2\!-\!\omega_{el}^2$.
Combining (\ref{ah})--(\ref{aj}), the NRCSR takes the compact form

	\begin{equation}S_{NR}({\bf q})=Z+C_{NR}({\bf q})
	-{1\over Z}\ {|F_{el}({\bf q})|^2\over
	G_{E,p}^2(Q_{el}^2)}.\label{ak}\end{equation}

\noindent The function $C_{NR}({\bf q})$ is simply the Fourier transform of
the spatial two-proton correlation function.  It has been defined so that
$C_{NR}({\bf q})\!=\!0$ for a system of $Z$ uncorrelated protons.  Since
$F_{el}({\bf q})$ is measured by elastic $(e,e^\prime)$ scattering, the
NRCSR (\ref{ak}) provides a direct measure of proton-proton correlations.
Since both $F_{el}({\bf q})$ and $C_{NR}({\bf q})$ are expected to
vanish in the limit of large $|{\bf q}|$, the approach of $S_{NR}({\bf q})
\!\rightarrow\!Z$ with increasing $|{\bf q}|$ is a test of the adequacy
of the theoretical assumptions.

Experimental studies of $(e,e^\prime)$ carried out in the momentum transfer
region up to $|{\bf q}|\!=\!550{\rm\ MeV/c}$ over the last decade have left
serious questions about the status of the NRCSR.  For light targets:
$A\!=2\!-\!4$, the measured Coulomb sum $S_{NR}({\bf q})\!\rightarrow\!Z$
for the largest momentum values reached.\cite{xa}  For heavier targets,
however, $S_{NR}\!<\!Z$ by 20-30\% compared to theoretical predictions based
on independent particle models,[3--8] with larger suppressions occurring
for larger nuclei.  This is despite the fact that Pauli correlations
are expected to vanish for $|{\bf q}|\!>\!2p_F\!\simeq\!550{\rm\ MeV/c}$.  We
shall not discuss all the possible effects which could be involved here, and
which have been reviewed elsewhere.\cite{pand,pandb}  However, one problem
which is always present in the application of the NRCSR to data is whether
the sum defined in (\ref{ae}) is in fact saturated by the experimental data:
not only is there an absolute relativistic limit $\omega\!<\!|{\bf q}|$ due to
the spacelike nature of virtual photons in electron scattering, but other
experimental difficulties prevent one from reaching even that limit in
practice.  We expect that extending experiments to higher energies and
momenta, such as will be done at CEBAF, may improve this situation.

The main problem dealt with in this paper is the extension of the Coulomb
sum rule to include the relativity of nucleons in inelastic scattering.
This becomes unavoidable for momentum transfers $|{\bf q}|\!\sim\!M$, the
nucleon mass, even if the nuclear target is essentially nonrelativistic.
This follows because even if the nucleons in a typical nucleus move at
essentially nonrelativistic velocities, they become highly relativistic after
the absorption of a virtual photon carrying large three-momentum.  Thus to
obtain an adequate description, one must describe nucleons with relativistic
wavefunctions, e.g., solutions to the Dirac equation.  This in turn
introduces the difficulty of treating antinucleon degrees of freedom, which
we deal with approximately.  Other issues also arise at higher energies,
including the composite nature of nucleons themselves.  This includes both
nucleon electromagnetic form factors, which we include, and inelastic
excitations of nucleons, which we do not.  We also do not consider the
effects of exchange currents among nucleons, e.g., as carried by charged
mesons.  These have been studied by Schiavilla {\it et al.},\cite{schi}
for example.  For the purpose of this paper, the nucleus consists of
interacting relativistic nucleons, considered to be elementary particles
with form factors.

A number of authors have considered relativistic extensions of the NRCSR
with the approximations just mentioned.  Walecka\cite{wal} made the most
direct extension by integrating the relativistic Coulomb response function
over all energies to allow the use of closure.  This results in a sum rule
which is nearly identical in form to the NRCSR.  Matsui\cite{mat} showed,
however, that such a sum rule would never be saturated by electron scattering,
for which $\omega\!<\!|{\bf q}|$.  He showed that, in the Fermi gas model,
excitations of the Fermi sea ($N$ scattering) are found entirely in the
spacelike response, while excitations of the Dirac sea ($N{\bar N}$
pair-production) are found entirely in the timelike response.  Unlike the
NRCSR, for a system of $Z$ Dirac protons the spacelike relativistic Coulomb sum
tends to $Z/2$ in the high-$|{\bf q}|$ limit as a result of this separation.
This work initiated interest in spacelike Coulomb sum rules based on energy
transfers of $\omega\!<\!|{\bf q}|$ only.  Several authors have studied the
case of a Fermi gas target.[13--16]  We will see that the exclusion of
timelike contributions from the Coulomb sum introduces kinematic factors
related to relativistic nucleon recoil which must be accounted for in the
formulation of {\it sum rules\/} for large $|{\bf q}|$.

DeForest\cite{defor} recognized the issue of relativistic nucleon
recoil and proposed a correction in the form of a modified electric
form factor: ${\bar G_E}(Q^2)\equiv G_E(Q^2)\sqrt{(1\!+\!\tau)/(1\!
+\!2\tau)}$, where $\tau\!\equiv\!Q^2/4M^2$, which would be factored from
the response function in the definition of the relativistic Coulomb sum.
He conjectured that the resulting sum rule should have an interpretation
similar to the NRCSR, but did not show this.  DeForest's prescription
has been used in the analysis of recent data.\cite{mez}  However, we will
show (in Sec.~5) that a result similar to his can be derived (RCSR-I) and
does lead to a sum rule which bears close resemblence similar to the NRCSR,
but is only accurate for Dirac nucleons, i.e., nucleons without anomalous
magnetic moments.  Donnelly {\it et al.}\cite{donn} have also proposed a
modification to the definition of the spacelike Coulomb sum.  By requiring
that their modified Coulomb sum tend to $Z$ in the large-$|{\bf q}|$
limit, an integral equation for the modified form factor can be derived,
then solved via an expansion in moments of the nucleon momentum.  In
lowest-order, this corresponds to the DeForest prescription evaluated
at zero nucleon momentum.  However, in higher order, this method
introduces an energy dependence into the Coulomb sum function which then
is {\it not\/} equivalent to a {\it non-energy-weighted\/} sum rule of the
NRCSR form, from which correlation information can be extracted.  We
discuss these issues further in the concluding section.

In this paper we derive a spacelike Coulomb sum rule valid for arbitrary
three-momentum transfers, in which the relativity of the nucleons is taken
into account.  We assume that nucleons are the only degrees of freedom
contributing to the spacelike sum and ignore, for example,
antinucleon degrees of freedom, meson exchange currents and internal
excitations of nucleons.  In Section 2, we review the formalism for electron
scattering from nuclear targets in the first-order Born approximation,
including nucleon form factors with an assumption about their off-shell
continuation.   In Section 3, we consider elastic scattering from a single,
free nucleon at rest as the simplest example of relativistic nucleon recoil
effects in the constant-{\bf q} response; this is shown to become a large
effect for $|{\bf q}|\!\sim\!M$.  In Section 4, we develop our
``nucleons-only'' approximation and derive a relativistic Coulomb sum
rule which is exact in its treatment of nucleon recoil and Fermi motion.
We give an explicit form for the two-body correlation function.  In
Section 5, we derive an expansion of the one-body term in moments of the
nucleon momentum.  This leads to practical sum rules for the analysis of
experimental data.  In Section 6, we present numerical results, including
a test of our moment expansion in the Fermi gas model.  Section 7 contains
a discussion and conclusions, including a recommendation for the application
of these sum rules to data analysis.

\setcounter{equation}{0}
\seceqb
\section{Formalism for $(e,e^\prime)$ Scattering}

We begin with the first-order Born approximation for the inelastic scattering
of ultrarelativistic $(|{\bf k}|\!>\!\!>\!m_e)$ electrons from nuclear
targets.  Initial and final electron four-momenta are denoted by $k^\mu\!
=\!(E,{\bf k})$ and ${k^\prime}^\mu\!=\!(E^\prime,{\bf k}^\prime)$, and initial
and final target four-momenta are denoted by $P^\mu\!=\!(P^0,{\bf P})$
and ${P^\prime}^\mu\!=\!({P^\prime}^0,{\bf P}^\prime)$, respectively.
The differential cross-section can be written in factored form:

	\begin{equation}{d^2\sigma\over d\Omega^\prime dE^\prime}=
	{\alpha^2\over q^4}{|{\bf k}^\prime|\over|{\bf k}|}
	L_e^{\mu\nu}(k^\prime,k)W_{\mu\nu}(P^\prime,P),
	\label{ba}\end{equation}

\noindent
where $q^\mu=(\omega,{\bf q})\equiv k^\mu-{k^\prime}^\mu={P^\prime}^\mu
-P^\mu$ is the four-momentum transferred to the target in the laboratory
frame by a single virtual photon, and $\alpha$ is the fine-structure
constant.  The lepton current tensor for an unpolarized electron beam is:

	\begin{equation}L_e^{\mu\nu}(k^\prime,k)
	\equiv{1\over2}\ \sum_{s s^\prime}
	\Bigl[{\bar u}_{s^\prime}({\bf k}^\prime)\gamma^\mu
	u_s({\bf k})\Bigr]\Bigl[{\bar u}_{s^\prime}({\bf k}^\prime)
	\gamma^\nu u_s({\bf k})\Bigr]^*,\label{bb}\end{equation}

\noindent where the fermion spinors $u_s({\bf k})$ are given in (\ref{aac}).

All of the interesting target physics is contained in the nuclear response
function:

	\begin{equation}W_{\mu\nu}(P,q)=\sum_f
	\langle f|{\hat J}_\mu(q)|i\rangle
	\langle f|{\hat J}_\nu(q)|i\rangle^*
	\delta(\omega-E_f+E_i),\label{bc}\end{equation}

\noindent where $|i\rangle$ and $|f\rangle$ denote initial\footnote
{Here, as in Section 1, we assume a nondegenerate ground state
$|i\rangle$.} and final nuclear many-body states, respectively. The tensor
$W_{\mu\nu}(P,q)$ may be decomposed into longitudinal and transverse
response functions, with respect to the polarization of the virtual photon
in a given frame: normally the target rest-frame.  (We will not discuss
the transverse response further in this paper.)  If the z-axis is chosen
along the three-momentum transfer {\bf q}, i.e., $q^\mu\!=\!(\omega,0,0,
|{\bf q}|)$, then the longitudinal polarization vector, chosen to obey
$e_L\!\cdot\!q\!=\!0$ (Lorentz gauge) and $e_L^2\!=\!1$, can be written
$e_L^\mu\!=\!(|{\bf q}|,0,0,\omega)/\sqrt{Q^2}$.  The longitudinal response
function is then defined

	\begin{equation}W_L(\omega,{\bf q})\equiv{e_L^\mu}^*
	W_{\mu\nu}e_L^\nu.\label{bd}\end{equation}

\noindent For theoretical reasons, it will be more convenient to work
in the Coulomb gauge, defining the Coulomb response function

	\begin{equation}W_C(\omega,{\bf q})\equiv W_{00}
	(\omega,{\bf q})={{\bf q}^2\over Q^2}\ W_L(\omega,{\bf q}),
	\label{be}\end{equation}

\noindent where the second relation reflects this change of gauge.
It is the Coulomb (not longitudinal) response function which appears
in (\ref{aa}).

In general, the current operator ${\hat J}_\mu(q)$ includes contributions
from both charged nucleons and charged mesons in the target ground state,
as discussed in the introduction.  However, for the present work we ignore
contributions to (\ref{bc}) from meson currents.  This allows us to write the
nucleon current operator in the one-body form:

	\begin{equation}{\hat J}_\mu(q)\equiv\int d^3x\
	{\rm e}^{i{\bf q\cdot x}}\ {\bar{\hat\psi}}({\bf x})
	\Gamma_\mu{\hat\psi}({\bf x}),\label{bf}\end{equation}

\noindent where ${\hat\psi}({\bf x})$ is the Dirac nucleon field operator
in the Schr\"odinger picture.  In general, the field operators and therefore
the current operator (\ref{bf}) involve both nucleon and antinucleon degrees of
freedom, as well as a sum over isospin projections.

In the relativistic impulse approximation, the field operators in (\ref{bf})
refer to free particles.  It is then natural to expand these field operators
in terms of free Dirac wavefunctions, as in (\ref{aab}).  The $\gamma
NN$ vertex operator $\Gamma_\mu$ may then be specified in terms of its
matrix elements between Dirac plane-wave spinors: ${\bar u}_{s^\prime}
({\bf p}^\prime)\Gamma_\mu u_s({\bf p})$, therefore it is conventional
to express $\Gamma_\mu$ in terms of Dirac matrices:

	\begin{equation}\Gamma_\mu=F_1\gamma_\mu+i{\kappa\over2M}
	F_2\sigma_{\mu\nu}q^\nu,\label{bg}\end{equation}

\noindent where $\kappa$ is the nucleon anomalous magnetic moment and $M$
is the nucleon mass.  In many-body calculations involving both protons
and neutrons, (\ref{bg}) includes an implicit sum over proton and neutron
isospin projections.  In general, the form factors $F_1$ and $F_2$ are
scalar functions of $p$, $p^\prime$ and $q$.

For scattering from free nucleons, the form factors $F_1$ and $F_2$ can
be shown to be functions only of the scalar variable $Q^2\!=\!{\bf q}^2\!
-\!\omega^2$.  These form factors are obtained from elastic $eN$ scattering
data, usually in terms of the more convenient electric and magnetic form
factors:

	\begin{eqnarray}G_E(Q^2)&\equiv&F_1(Q^2)-\kappa\tau F_2(Q^2)
	\nonumber\\ G_M(Q^2)&\equiv&F_1(Q^2)+\kappa F_2(Q^2)
	\label{bh}\end{eqnarray}

\noindent respectively, where $\tau\!\equiv\!Q^2/4M^2$.  Then $F_1$
and $F_2$ are given by

	\begin{eqnarray}F_1(Q^2)&=&{G_E(Q^2)+\tau G_M(Q^2)\over1+\tau}
	\nonumber\\
	& &\nonumber\\
	F_2(Q^2)&=&{G_M(Q^2)-G_E(Q^2)\over1+\tau}
	\label{bi}\end{eqnarray}

\noindent with the normalization convention that $F_1^p(0)\!=\!F_2^p(0)
\!=\!F_2^n(0)\!=\!1$ and $F_1^n(0)\!=\!0$.

The continuation of these form factors for interacting nucleons is
not unique, requiring dynamical information not contained in the
impulse approximation.  A common assumption is to use the on-shell
values $F_1(Q^2)$ and $F_2(Q^2)$ in constructing the vertex operator
(\ref{bg}).  An equally reasonable assumption, which we adopt in this paper,
is to evaluate {\it only\/} the electric and magnetic form factors at
the off-shell variable $Q^2$, i.e., the off-shell $F_1$ and $F_2$ are
defined:

	\begin{eqnarray}F_1(Q^2,{\tilde Q}^2)&\equiv&{G_E(Q^2)
	+{\tilde\tau}G_M(Q^2)\over1+{\tilde \tau}}\nonumber\\
	& &\nonumber\\ F_2(Q^2,{\tilde Q}^2)&\equiv&{G_M(Q^2)-G_E(Q^2)
	\over1+{\tilde\tau}}\label{bj}\end{eqnarray}

\noindent where ${\tilde\tau}\!\equiv\!{\tilde Q}^2/4M^2$ and ${\tilde Q}^2
\!\equiv\!{\bf q}^2\!-\!(E_{\bf p+q}\!-\!E_{\bf p})^2$.  Here $E_{\bf p}
\equiv\sqrt{{\bf p}^2+M^2}$ is the energy of a free nucleon with
three-momentum ${\bf p}$.

The crucial feature of the off-shell prescription (\ref{bj}) is that the photon
energy $\omega$ enters $\Gamma_\mu$ in (\ref{bg}) only through the form factors
$G_E(Q^2)$ and $G_M(Q^2)$.  Consider the limit of point nucleons, for which
the form factors $F_1(Q^2)$ and $F_2(Q^2)$ in (\ref{bg}) become constants:
then the plane-wave matrix elements ${\bar u}_{s^\prime}({\bf p}\!+\!{\bf q})
\Gamma_\mu u_s({\bf p})$ depend only on the three-vectors {\bf p} and
{\bf q}.  For extended particles, both the $F$'s and the $G$'s depend on
the scalar $Q^2$, and may also depend on another scalar, which we take to
be ${\tilde Q}^2$.  (This choice is not unique, but is sensible given that
{\bf p} and {\bf q} are the only other variables at the $\gamma NN$ vertex
in the impulse approximation.)  One may assume that {\it either\/} the $F$'s
or the $G$'s, but not both sets, are functions of $Q^2$ only; we have made
the second choice in (\ref{bj}).  This prescription is necessary to arrive at
the particularly simple forms for the relativistic Coulomb sum rule which
follows.

We adopt the following parameterizations\cite{pb} for the nucleon
electric and magnetic form factors appearing in (\ref{bj}):
\vfill\eject

	\begin{eqnarray}
	G_{E,p}(Q^2)&=&(1+Q^2/.71{\rm GeV}^2)^{-2}\label{bk}\\
	G_{M,p}(Q^2)&=&(1+\kappa_p)G_{E,p}(Q^2)\label{bl}\\
	G_{M,n}(Q^2)&=&\kappa_n G_{E,p}(Q^2)\label{bm}\\
	G_{E,n}(Q^2)&=&0\label{bn}
	\end{eqnarray}

\noindent where $\kappa_p\!=\!+1.79$ and $\kappa_n\!=\!-1.91$ are
the proton and neutron anomalous magnetic moments, respectively.
Although the neutron electric form factor $G_{E,n}(Q^2)$ is not
identically zero in Ref.~{}\cite{pb}, it is known to be small compared
to $G_{E,p}(Q^2)$.  Other nonzero forms for $G_{E,n}(Q^2)$ have also
been given,\cite{pb,hoh} and could be adopted for use in the present
work.  However, we shall make use of the fact that {\it all\/} form
factors have $G_{E,p}(Q^2)$ as a common factor to simplify the form of
the relativistic Coulomb sum rule to follow.  We note, however, that
this proportionality cannot be exact for nonzero $G_{E,n}(Q^2)$,
since $G_{E,n}(0)\!=\!0$.

\setcounter{equation}{0}
\seceqc
\section{Single-nucleon Coulomb Scattering}

In this section, we consider the simplest nuclear target, one free
nucleon, in the response function formalism of the previous section.
This allows us to illustrate directly some of the features of relativity
which also occur in the general nuclear case, and suggest how they
may be treated in the formulation of relativistic Coulomb sum rules.
In this case, the scattering is elastic on the target nucleon; the
differential cross-section in the laboratory frame, which may be obtained
from (\ref{ba}), is the Rosenbluth formula with form factors (\ref{bh}).
We are interested only in the Coulomb scattering, which is expressed
in terms of the Coulomb response function $W_C(\omega,{\bf q})\!\equiv
\!W_{00}(\omega,{\bf q})$.

For a free nucleon, the Coulomb response function (\ref{be}) may be evaluated
from (\ref{bc}) and (\ref{bf}) by specifying that initial and final nuclear
states are simply Dirac plane-wave states of momenta {\bf p} and ${\bf p}
^\prime\!=\!{\bf p}\!+\!{\bf q}$, respectively.  Using the nucleon field
operator of (\ref{aab}), we find:

	\begin{equation}W^N_C({\bf p},q)={L^N_{00}
	({\bf p},q)\over4E_{\bf p+q}E_{\bf p}}
	\ \delta(\omega\!-\!E_{\bf p+q}\!+\!E_{\bf p})
	,\label{ca}\end{equation}

\noindent where the 00-component of the nucleon current tensor is given
by

	\begin{equation}L^N_{00}({\bf p},q)
	\equiv{1\over2}\ \sum_{s s^\prime}
	\Bigl[{\bar u}_{s^\prime}({\bf p}^\prime)\Gamma_0
	u_s({\bf p})\Bigr]\Bigl[{\bar u}_{s^\prime}({\bf p}^\prime)
	\Gamma_0 u_s({\bf p})\Bigr]^*,\label{cb}\end{equation}

\noindent and the the energy factors in the denominator of (\ref{ca}) reflect
our convention that the fermion spinors (\ref{aac}) are normalized to $2E$
particles per unit volume.  The delta-function in (\ref{ca}) guarantees energy
conservation for the free nucleon states: $\omega\!=\!E_{\bf p+q}
\!-\!E_{\bf p}$.

Notice that only nucleon, i.e., not antinucleon, states appear in (\ref{cb}).
The energy required to excite a free particle of momentum {\bf p} into
a state with momentum ${\bf p}\!+\!{\bf q}$ satisfies $E_{\bf p+q}\!-
\!E_{\bf p}\!<\!|{\bf q}|$, hence the process $N\!\rightarrow\!N$
is restricted entirely to the spacelike region $\omega\!<\!|{\bf q}|$.
In contrast, the energy required to excite an $N{\bar N}$ pair from the
vacuum satisfies $\omega\!=\!E_{\bf p+q}\!+\!E_{\bf p}\!>\!|{\bf q}|$,
and hence is restricted entirely to the timelike region $\omega\!>\!
|{\bf q}|$.  (Pair-production arising due to scattering from a free
nucleon may be considered either an excitation of the nucleon itself
or a two-body, i.e., meson exchange, process and is ignored here.)
Thus the antinucleon components of the field operators of (\ref{aab}) do
not appear in the (spacelike) response function for a free nucleon.
We shall find it advantageous to use this separation of excitation
spectra to explicitly remove the contribution of $N{\bar N}$ pairs
from the relativistic sum rule. (This issue was first raised by
Matsui.\cite{mat})

Returning to the Coulomb response function, we first give an explicit
form for the 00-component of the nucleon current tensor (\ref{cb}).  After
substituting (\ref{bg}) and (\ref{bj}) into (\ref{cb}), a straightforward
trace calculation yields

	\begin{equation}L^N_{00}({\bf p},q)={G_E^2(Q^2)\over
	1+{\tilde\tau}}(E_{\bf p+q}\!+\!E_{\bf p})^2
	+{G_M^2(Q^2)\over1+{\tilde\tau}}\Bigl[
	{\tilde\tau}(E_{\bf p+q}\!+\!E_{\bf p})^2
	-(1+{\tilde\tau}){\bf q}^2\Bigr],\label{cc}\end{equation}

\noindent where, of course, for a free nucleon ${\tilde\tau}\!=\!\tau$
and ${\tilde Q}^2\!=\!Q^2$.  If we put the initial nucleon at rest,
evaluation of (\ref{cc}) at ${\bf p}\!=\!0$ yields

	\begin{equation}L^N_{00}({\bf p},q)\bigl\vert_{{\bf p}=0}
	=2M(E_{\bf q}+M)\ G_E^2(Q^2),\label{cd}\end{equation}

\noindent where we have used ${\tilde\tau}\!=\!(E_{\bf q}\!-\!M)/2M$
at ${\bf p}\!=\!0$.  Thus stationary nucleons couple {\it only\/} through
their electric form factor.  (A sensible feature of the off-shell
prescription (\ref{bj}) is that this reduction also occurs for interacting
nucleons.)  The response function is obtained by substituting (\ref{cd})
into (\ref{ca}).

The Coulomb sum function is obtained by integrating $W_C(\omega,{\bf q})/G
_{E,p}^2(Q^2)$ over energy transfer $\omega$, as in (\ref{ae}).  For
electron scattering from a free nucleon, it is natural to restrict
the integral to include only spacelike ($\omega\!<\!|{\bf q}|$) photons,
thus excluding $N{\bar N}$ pair production:

	\begin{equation}\int_{\omega_{el}^+}^{|{\bf q}|}d\omega\
	{W^N_C({\bf p},q)\over G_E^2(Q^2)}
	\biggr\vert_{{\bf p}=0}={E_{\bf q}+M
	\over2E_{\bf q}}
	={M\over E_{\bf q}}{{\bf q}^2\over Q^2}
	\biggl\vert_{\omega=E_{\bf q}-M}\label{ce}
	\end{equation}

\noindent where in the second relation we have made use of the nucleon
energy-momentum relation at ${\bf p}\!=\!0$ to express the result in
a form which lends itself easily to interpretation.  The first factor
$M/E_{\bf q}$ has its origin in the energy denominator $E_{\bf p+q}$ in
(\ref{ca}), which in turn comes from the relativistic normalization of the
fermion spinors (\ref{aac}).  The second factor ${\bf q}^2/Q^2$ arises
from our choice to work in Coulomb gauge, as can be seen by comparing
with (\ref{be}).  Taken together, these factors represent the effect of
relativistic recoil (in Coulomb gauge) for a single nucleon.  The first
relation in (\ref{ce}) shows that this recoil effect tends to unity in the
nonrelativistic limit $|{\bf q}|\!<\!\!<\!M$, and tends to the value $1/2$
in the ultrarelativistic limit $|{\bf q}|\!>\!\!>\!M$.  (It is in this
sense that Coulomb gauge is a convenient choice, since the analog of
(\ref{ce}) in Lorentz gauge tends to zero in the limit $|{\bf q}|
\!>\!\!>\!M$.)

Matsui\cite{mat} and DoDang {\it et al.}\cite{ddb} have previously noticed
this limiting value of $1/2$ for the spacelike Coulomb sum, for a Fermi
gas of Dirac ($G_E\!=\!G_M\!=\!1$) nucleons.  The Fermi gas result is
easily obtained by inserting (\ref{cc}) into (\ref{ca}) and integrating over
the Fermi sphere $|{\bf p}|\!<\!p_F$ with an appropriate factor to account
for Pauli blocking, as is done in (\ref{fb}).  It is easy to show
that this quantity approaches $Z/2$ (for $Z$ protons) as $|{\bf q}|\!
\rightarrow\!\infty$.  That this is a purely kinematic effect of
relativistic nucleon recoil is clear from the single-nucleon example
(\ref{ce}).

The high-$|{\bf q}|$ behavior of (\ref{ce}) is in contrast to that of the NRCSR
(\ref{ak}), for which $S_{NR}({\bf q})\!\rightarrow\!Z$ ($Z\!=\!1$ here) as
$|{\bf q}|\!\rightarrow\!\infty$, suggesting that in the relativistic
limit, the Coulomb sum does {\it not\/} directly count the number of
charged scatterers, as in the nonrelativistic case.  A simple modification
of (\ref{ce}) would, however, seem to remedy this for the case of a single
nucleon: divide both sides by the kinematic recoil factor on the
right-hand side, so that the large-$|{\bf q}|$ limit is now unity, as would
be expected for a sum rule.  That this actually corresponds to a sum rule
for the many-body case, with a suitable modification for the nucleon momentum
distribution (Fermi motion) in the target, is shown in the next two
sections.

A final note:  For a free nucleon, the delta-function of (\ref{ca}) appears in
all components of $W_{\mu\nu}$.  To obtain the Rosenbluth formula, one must
integrate (\ref{ba}) over $dE^\prime$ (or $d\omega$) at fixed electron
scattering angle $\theta$, rather than at fixed ${\bf q}$, as in the Coulomb
sum (\ref{ce}).  The constraint that $\omega$ and {\bf q} are {\it not\/}
independent variables, but are related by $Q^2\!=\!4EE^\prime{\rm sin}^2
(\theta/2)$, then gives rise to the familiar recoil factor:

	\begin{equation}{E^\prime\over E}=\biggl[1+{2E\over M}{\rm sin}^2
	\biggl({\theta\over2}\biggr)\biggr]^{-1}\label{cf}\end{equation}

\noindent which reduces to unity in the nonrelativistic limit $E\!<\!\!<
\!M$, where $E$ is the electron beam energy.  Thus it is well-known that
not only form factors, but also nucleon recoil factors, modify the
differential cross-section for electron scattering in the relativistic
domain.

\setcounter{equation}{0}
\seceqd
\section{Relativistic Coulomb Sum Rule for Interacting Nucleons}

In this section, we derive a sum rule for a target of interacting nucleons,
where the relativity of the nucleons is taken into account.  Following
the example of a single nucleon in the previous section, we find that
if we restrict the sum to spacelike four-momentum transfers, which
corresponds to $(e,e^\prime)$ scattering experiments, we are led to a
``nucleons-only'' approximation in which antinucleon degrees of freedom
are ignored.  The new feature in the sum rule given here is the kinematic
correction for relativistic recoil, which depends on the momentum of the
nucleons in the target.  In the following section, we show how this
correction can be calculated in successive approximations in terms of
moments of the nucleon momentum.

We begin with the Coulomb response function in the rest-frame of a many-body
nuclear system, which is given by $W_{00}(P,q)$ of (\ref{bc}), evaluated at
${\bf P}\!=\!0$:

	\begin{equation}W_C(\omega,{\bf q})=\sum_f
	|M_{fi}(\omega,{\bf q})|^2\
	\delta(\omega-E_f+E_i),\label{da}\end{equation}

\noindent where the Coulomb transition matrix element is:

	\begin{equation}M_{fi}(\omega,{\bf q})\equiv\langle f|
	{\hat J}_0(\omega,{\bf q})|i\rangle
	=\int d^3x\ {\rm e}^{i{\bf q\cdot x}}\
	\langle f|{\bar{\hat\psi}}({\bf x})\Gamma_0{\hat\psi}
	({\bf x})|i\rangle.\label{db}\end{equation}

\noindent These are the relativistic equivalents of (\ref{ac}) and (\ref{ad}).
The $\gamma NN$ vertex operator $\Gamma_0$ is given by (\ref{bg})
and includes an implicit sum over isospin projections, and the field
operator ${\hat\psi}({\bf x})$ is given in (\ref{aab}) and includes both
nucleon and antinucleon degrees of freedom.  As discussed in Section 2,
the nucleon form factors are assumed to be given by (\ref{bj}) such that
the only $\omega$ dependence in ${\hat J}_0(q)$ enters through the nucleon
electric and magnetic form factors, $G_E(Q^2)$ and $G_M(Q^2)$.  However,
given the specific functional forms in (\ref{bk})--(\ref{bn}), one can
instead consider {\it all\/} the $\omega$ dependence in (\ref{db}) to enter
only through the common form factor $G_{E,p}(Q^2)$.  Dividing (\ref{db})
by $G_{E,p}(Q^2)$ therefore removes {\it all\/} the $\omega$ dependence
from the nuclear matrix element $M_{fi}$.  Dividing (\ref{da}) by $G_{E,p}
^2(Q^2)$ then allows us to generate a {\it non-energy-weighted\/} sum
rule, as follows.

We consider the Coulomb sum function defined by

	\begin{equation}\Sigma({\bf q})\equiv\int_{\omega_{el}^+}
	^{|{\bf q}|}d\omega\ {W_C(\omega,{\bf q})\over G_{E,p}^2(Q^2)},
	\label{dc}\end{equation}

\noindent where the integration is performed only over energies $\omega\!
<\!|{\bf q}|$, corresponding to electron scattering experiments, as in
(\ref{ce}).  This is unlike the nonrelativistic Coulomb sum (\ref{ae}),
in which the integration is over all $\omega\!>\!\omega_{el}$ in order to
invoke closure.  Substituting (\ref{da}) into (\ref{dc}) and performing
the integration, we obtain:

	\begin{equation}\Sigma({\bf q})=\sum_{f\not=i}
	^{f:\omega<|{\bf q}|}\ \biggl|{M_{fi}(\omega,{\bf q})
	\over G_{E,p}(Q^2)}\biggr|^2,\label{dd}\end{equation}

\noindent where the sum over all final states in (\ref{da}) is now reduced
to a sum over only those final states $f\!\not=\!i$ which are accessed by
spacelike four-momentum transfers.  For the case of a free nucleon target
at rest, (\ref{dd}) reduces to (\ref{ce}).

The nucleon field operator (\ref{aab}), and therefore the transition matrix
elements (\ref{db}), involve both nucleon and antinucleon degrees of
freedom. However, because electron scattering experiments produce spacelike
virtual photons exclusively, the resulting nuclear excitations are
restricted to those whose excitation energies satisfy $\omega\!=\!E_f\!
-\!E_i\!<\!|{\bf q}|$, as indicated in (\ref{dd}).  We have already seen
in (\ref{ce}), for the case of a free nucleon, that restricting the energy
transfer $\omega$ removes the contribution of $N{\bar N}$ pair production
from the Coulomb sum, which would enter if $\omega\!>\!|{\bf q}|$ were
included in (\ref{dc}).  Similarly, for a noninteracting Fermi gas,
only excitations involving $a^\dagger_{{\bf p}^\prime s^\prime}
a_{{\bf p}s}$ ($N$ scattering) enter into the spacelike sum.
This is because there are no antinucleons present in the
renormalized nuclear ground state, thus the terms involving $b^\dagger
_{{\bf p}s}b_{{\bf p}^\prime s^\prime}$ (${\bar N}$ scattering) and
$b_{{\bf p}^\prime s^\prime}a_{{\bf p}s}$ ($N{\bar N}$ pair-annihilation)
vanish identically, and because the terms involving $a^\dagger_{{\bf p}
^\prime s^\prime}b^\dagger_{{\bf p}s}$ ($N{\bar N}$ pair-creation) are
restricted entirely to the timelike region, since $\omega\!=\!E_{\bf p+q}
\!+\!E_{\bf p}\!>\!|{\bf q}|$.  Furthermore, the separation of nucleon
and antinucleon degrees of freedom is complete in the Fermi gas model:
not only are antinucleon scattering and $N{\bar N}$ pair terms completely
removed from the spacelike sum, but the $N\!\rightarrow\!N$ contribution
is guaranteed to lie completely in the spacelike region through the
relation $\omega\!=\!E_{\bf p+q}\!-\!E_{\bf p}\!<\!|{\bf q}|$.  (This
feature is unaffected by the inclusion of scalar and vector potentials
like those in Quantum Hadrodynamics, for example, as noted by Do~Dang
{\it et al.}\cite{ddb})  Thus, in the Fermi gas model, the spacelike
nuclear response (\ref{dd}) can be represented by effectively dropping
all reference to antinucleon degrees of freedom at the level of the
field operator (\ref{aab}).

For an interacting nuclear system, however, the separation of nucleon
and antinucleon degrees of freedom is no longer exact.  In general, both
$N$ and ${\bar N}$ degrees of freedom can contribute to the spacelike
nuclear response, and some of the nucleons-only response $N\!\rightarrow
\!N$, which is entirely spacelike in the Fermi gas model, can be pushed
into the timelike region by two-particle interactions.  However, we can
assume that true $N{\bar N}$ pair production will still occur mostly
for $\omega\!>\!|{\bf q}|$, and therefore can be neglected in the Coulomb
sum (\ref{dd}).  Although tests of this assumption are outside the scope
of this paper, we note that it is satisfied exactly for infinite nuclear
matter in the RPA, i.e., a spacelike photon excites only spacelike
excitations even through the summation of an infinite number of RPA
ring diagrams, and does not produce $N{\bar N}$ pairs.  Antinucleon
contributions, i.e., those involving $b^\dagger_{{\bf p}s}b_{{\bf p}
^\prime s^\prime}$ and $b_{{\bf p}^\prime s^\prime}a_{{\bf p}s}$,
enter as relativistic effects in the nuclear target structure.  In
Hartree calculations for finite nuclei, these terms make only small
($\sim\!1\%$) contributions to the relevant (vector) density,\cite{vac}
and shall also be neglected in this paper.  Therefore, we adopt a
``nucleons-only'' approximation, in which antinucleon degrees of freedom
are removed from discussion at the level of the field operator (\ref{aab}).
In effect, we are assuming that the {\it dominant\/} relativistic effect
comes from single-nucleon recoil at high $|{\bf q}|$, and not from
relativistic aspects of nuclear structure.  Thus the initial and final
nuclear states in (\ref{dd}) will be treated as interacting many-nucleon
states without ${\bar N}$ components.

Substituting the field operators given by (\ref{aab}) into the transition
matrix element (\ref{db}), we have in our ``nucleons-only'' approximation:

	\begin{equation}M_{fi}(\omega,{\bf q})\simeq G_{E,p}(Q^2)
	\sum_{{\bf p}\sigma}\sum_{ss^\prime}
	j_{s^\prime s,\sigma}({\bf p},{\bf q})
	\langle f|a_{{\bf p+q}s^\prime\sigma}^\dagger
	a_{{\bf p}s\sigma}|i\rangle,\label{de}\end{equation}

\noindent where the ``reduced'' spinor matrix element is defined:

	\begin{equation}j_{s^\prime s,\sigma}({\bf p},{\bf q})\equiv
	{{\bar u}_{s^\prime\sigma}({\bf p}\!+\!{\bf q})
	\over\sqrt{2E_{\bf p+q}}}{\Gamma_0({\bf p},q)\over G_{E,p}(Q^2)}
	{u_{s\sigma}({\bf p})\over\sqrt{2E_{\bf p}}}.\label{df}\end{equation}

\noindent In (\ref{de}) and (\ref{df}), nucleon spin states are represented by
Latin indices and nucleon isospin states are represented by Greek indices.
The operator $\Gamma_0({\bf p},q)$ is diagonal in isospin space, therefore
the transition matrix element (\ref{de}) simply involves a sum over proton and
neutron isospin projections.  The term ``reduced'' in (\ref{df}) refers to the
division by $G_{E,p}(Q^2)$, following (\ref{dc}). The matrix element (\ref{df})
is then a function only of {\bf p} and {\bf q}, i.e., not of $\omega$, through
our assumption of proportional form factors.  Substituting (\ref{de}) into
(\ref{dd}) leads to:

	\begin{eqnarray}\Sigma({\bf q})&=&\sum_{{\bf p}{\bf p}^\prime}
	\sum_{rr^\prime}\sum_{ss^\prime}\sum_{\sigma\rho}
	\ j^*_{r^\prime r,\rho}({\bf p}^\prime,{\bf q})
	\ j_{s^\prime s,\sigma}({\bf p},{\bf q})\nonumber\\
	&&\nonumber\\ &&
	\Biggl\lbrace\ \sum_{f:\omega<|{\bf q}|}
	\langle i|a^\dagger_{{\bf p}^\prime r\rho}
	a_{{\bf p}^\prime+{\bf q} r^\prime\rho}|f\rangle
	\langle f|a^\dagger_{{\bf p+q}s^\prime\sigma}
	a_{{\bf p}s\sigma}|i\rangle\nonumber\\
	&&\qquad\qquad-\langle i|a^\dagger_{{\bf p}^\prime r\rho}
	a_{{\bf p}^\prime+{\bf q} r^\prime\rho}|i\rangle
	\langle i|a^\dagger_{{\bf p+q}s^\prime\sigma}
	a_{{\bf p}s\sigma}|i\rangle\Biggr\rbrace,
	\label{dg}\end{eqnarray}

\noindent where we have subtracted the elastic term $f\!=\!i$ explicitly,
in order to obtain a sum over the complete set of nuclear final states. For
notational simplicity, we assume a nondegenerate ground state $|i\rangle$.
Equation (\ref{dg}) is then the most general form for the Coulomb sum function
(\ref{dc}), given our nucleons-only approximation and our assumptions about
nucleon form factors and their off-shell continuation.

In order to obtain a sum rule from (\ref{dg}), we must sum over a complete set
of final states $|f\rangle$ and use closure.  The completeness of such a
set necessarily includes excited nuclear states accessed by {\it both\/}
spacelike ($\omega\!<\!|{\bf q}|$) and timelike ($\omega\!>\!|{\bf q}|$)
photons, which can be expressed

	\begin{equation}\sum_{f:\omega<|{\bf q}|}|f\rangle\langle f|
	+\sum_{f:\omega>|{\bf q}|}|f\rangle\langle f|=1.\label{dh}
	\end{equation}

\noindent It is consistent with the nucleons-only approximation imposed in
(\ref{de}) to neglect the contribution of timelike states $f:\!\omega\!>\!|
{\bf q}|$ to the closure of the sum in (\ref{dg}),
since these states predominantly involve
antinucleons.  For a uniform Fermi gas in the Hartree-Fock approximation,
we have already seen that the separation between $N$ and ${\bar N}$ degrees
of freedom guarantees that the spacelike Coulomb sum be completely exhausted
in the nucleons-only approximation.  Similarly, for an interacting Fermi gas
in the RPA, nucleons are excited {\it only\/} for $\omega\!<\!|{\bf q}|$, and
$N{\bar N}$ pairs are produced {\it only\/} for $\omega\!>\!|{\bf q}|$, thus
the same argument holds.  It is possible, however, that restricting the sum
in (\ref{dd}) to spacelike states $f\!:\!\omega\!<\!|{\bf q}|$ does {\it not\/}
in general exhaust the nucleons-only excitation spectrum, e.g., finite size
effects and two-body interactions beyond the RPA may push some of the
nucleons-only response into the timelike region.  We are aware of no general
argument which applies.  We will simply assume that the nucleons-only
excitation spectrum {\it is\/} saturated by states $f\!:\!\omega\!<\!|{\bf
q}|$, so that we can invoke closure in (\ref{dg}).  Extension of the sum to
include nucleonic excitations with $\omega\!>\!|{\bf q}|$ could be done by
theoretical means, as in Schiavilla {\it et al.},\cite{spf} for example.
Applying closure to the first term in (\ref{dg}) removes the
sum over final states, and the expression in curly brackets becomes:

	\begin{eqnarray}\biggl\lbrace...\biggr\rbrace
	&\rightarrow&\biggl\lbrace
	\langle i|a^\dagger_{{\bf p}^\prime r\rho}
	a_{{\bf p}^\prime+{\bf q} r^\prime\rho}
	a^\dagger_{{\bf p+q}s^\prime\sigma}
	a_{{\bf p}s\sigma}|i\rangle\nonumber\\
	&&\nonumber\\
	&-&\langle i|a^\dagger_{{\bf p}^\prime r\rho}
	a_{{\bf p}^\prime+{\bf q} r^\prime\rho}
	|i\rangle\langle i|a^\dagger_{{\bf p+q}s^\prime\sigma}
	a_{{\bf p}s\sigma}|i\rangle\biggr\rbrace.
	\label{di}\end{eqnarray}

\noindent Equation (\ref{dg}) is now a {\it sum rule}, equating the sum
(\ref{dc}) to the ground state expectation values in (\ref{di}).

To cast the sum rule in a more transparent form, we separate out the
one- and two-body parts of the operator in the first term of (\ref{di}),
using the momentum-space anticommutation relations for the nucleon
creation and destruction operators:

	\begin{equation}\bigl\lbrace a_{{\bf p}s\sigma},
	a^\dagger_{{\bf p}^\prime s^\prime\sigma^\prime}
	\bigr\rbrace=\delta_{{\bf p}{\bf p}^\prime}
	\delta_{s s^\prime}\delta_{\sigma\sigma^\prime}
	,\label{dj}\end{equation}

\noindent where the anticommutator of any two creation or destruction
operators vanishes.  Moving all creation operators to the left in the
first term of (\ref{di}), we obtain:

	\begin{eqnarray}\biggl\lbrace...\biggr\rbrace
	&\rightarrow&\biggl\lbrace\langle i|a^\dagger_{{\bf p}s\sigma}
	a_{{\bf p}s\sigma}|i\rangle\delta_{{\bf pp}^\prime}
	\delta_{r^\prime s^\prime}\delta_{\rho\sigma}
	\delta_{rs}\nonumber\\
	&&\nonumber\\
	&+&\langle i|a^\dagger_{{\bf p+q}s^\prime\sigma}
	a^\dagger_{{\bf p}^\prime r\rho}
	a_{{\bf p}^\prime+{\bf q} r^\prime\rho}
	a_{{\bf p}s\sigma}|i\rangle\nonumber\\
	&&\nonumber\\
	&-&\langle i|a^\dagger_{{\bf p}^\prime r\rho}
	a_{{\bf p}^\prime+{\bf q} r^\prime\rho}
	|i\rangle\langle i|a^\dagger_{{\bf p+q}s^\prime\sigma}
	a_{{\bf p}s\sigma}|i\rangle\biggr\rbrace,
	\label{dk}\end{eqnarray}

\noindent where we have extracted $\delta_{rs}$ from the first matrix
element, which is diagonal in spin projection.  Inserting (\ref{dk}) into
(\ref{dg}) allows us to separate the sum rule expression into one-body,
two-body and elastic amplitudes, in complete analogy to the
nonrelativistic expression (\ref{ah}).  Here, however, we have used
momentum operators, rather than local field operators, in order
to eliminate the antinucleon components.

We separate the sum rule as follows:

	\begin{equation}\Sigma({\bf q})\equiv\Sigma^{(1)}({\bf q})
	+\Sigma^{(2)}({\bf q}).\label{dl}\end{equation}

\noindent The first term gives the one-body contribution:

	\begin{equation}\Sigma^{(1)}({\bf q})=2\sum_{{\bf p}\sigma}
	n_\sigma({\bf p})\ r_\sigma({\bf p},{\bf q}),\label{dm}
	\end{equation}

\noindent where we have defined the momentum distribution function
$n_\sigma({\bf p})\!\equiv\!\langle i|a^\dagger_{{\bf p}s\sigma}
a_{{\bf p}s\sigma}|i\rangle$ for isospin projection $\sigma$ (which
we assume is independent of $s$), and the relativistic recoil factor

	\begin{eqnarray}r_\sigma({\bf p},{\bf q})
	&\equiv&{1\over2}\sum_{ss^\prime}\bigl\vert
	j_{s^\prime s,\sigma}({\bf p},{\bf q})\bigr\vert^2
	\nonumber\\
	&&\nonumber\\
	&=&{1\over G^2_{E,p}(Q^2)}{L_{00}^\sigma({\bf p},q)
	\over4E_{\bf p+q}E_{\bf p}},\label{dn}\end{eqnarray}

\noindent where $L_{00}^\sigma({\bf p},q)$ is the nucleon tensor
component (\ref{cc}).  In the nonrelativistic limit $|{\bf q}|\!\ll\!M$,
$r_\sigma({\bf p},{\bf q})\!\rightarrow\!1$ and $\Sigma^{(1)}({\bf q})\!
\rightarrow\!Z$, as in (\ref{ah}).  For relativistic momenta $|{\bf q}|\!
\gg\!M$, however, we have $r_\sigma({\bf p},{\bf q})\!<\!1$, which
represents the kinematic effect of relativistic nucleon recoil, as in
(\ref{ce}).  If the Fermi momentum in (\ref{dm}) could be neglected
entirely, i.e., if we were to set ${\bf p}\!=\!0$ in (\ref{dn}), then
$\Sigma^{(1)}({\bf q})/Z$ would be given by (\ref{ce}).  The second term
of (\ref{dl}) is given by

	\begin{eqnarray}\Sigma^{(2)}({\bf q})&=&
	\sum_{{\bf p}{\bf p}^\prime}\sum_{s s^\prime}
	\sum_{r r^\prime}\sum_{\sigma\rho}
	\ j^*_{r^\prime r,\rho}({\bf p}^\prime,{\bf q})
	\ j_{s^\prime s,\sigma}({\bf p},{\bf q})\nonumber\\
	&&\nonumber\\
	&&\quad\Bigl\lbrace
	\langle i|a^\dagger_{{\bf p+q}s^\prime\sigma}
	a^\dagger_{{\bf p}^\prime r\rho}
	a_{{\bf p}^\prime+{\bf q} r^\prime\rho}
	a_{{\bf p}s\sigma}|i\rangle\nonumber\\
	&&\nonumber\\
	&&\qquad-\langle i|a^\dagger_{{\bf p}^\prime r\rho}
	a_{{\bf p}^\prime+{\bf q} r^\prime\rho}
	|i\rangle\langle i|a^\dagger_{{\bf p+q}s^\prime\sigma}
	a_{{\bf p}s\sigma}|i\rangle\Bigr\rbrace.
	\label{do}\end{eqnarray}

\noindent
This expression contains the full two-body probability in momentum space,
in analogy with the second term of (\ref{ah}), but is weighted by the
kinematic factor $j^*_{r^\prime r,\rho}({\bf p}^\prime,{\bf q})\
j_{s^\prime s,\sigma}({\bf p},{\bf q})$.

To separate out the correlation part of $\Sigma^{(2)}({\bf q})$ as in
(\ref{ai}), we first evaluate (\ref{do}) for an uncorrelated target ground
state, for which we obtain

	\begin{equation}\Sigma_{un}^{(2)}({\bf q})=-\sum_\sigma
	{1\over N_\sigma}\Bigl\vert\sum_{\bf p}\sum_{ss^\prime}
	\ j_{s^\prime s,\sigma}({\bf p},{\bf q})\langle i|
	a^\dagger_{{\bf p+q}s^\prime\sigma}a_{{\bf p}s\sigma}
	|i\rangle\Bigr\vert^2,\label{dp}\end{equation}

\noindent where $N_\sigma\!=\!Z,N$ for protons and neutrons, respectively.
(In the uncorrelated case, there is no contribution in (\ref{dp}) from
$\sigma\!\not=\!\rho$.)  We may then define the correlation function

	\begin{equation}C({\bf q})\equiv\Sigma^{(2)}({\bf q})
	-\Sigma_{un}^{(2)}({\bf q})\label{dq}\end{equation}

\noindent to give a measure of true two-body correlations, and
rewrite (\ref{dl})

	\begin{equation}\Sigma({\bf q})=\Sigma^{(1)}({\bf q})
	+C({\bf q})+\Sigma_{un}^{(2)}({\bf q}),\label{dr}
	\end{equation}

\noindent in analogy with (\ref{ak}).  This result will be referred to as
the relativistic Coulomb sum rule (RCSR).

The uncorrelated term $\Sigma_{un}^{(2)}({\bf q})$ is related to the
square of the elastic target form factor, with the difference that
the proton and neutron contributions are added {\it incoherently\/} in
(\ref{dp}).  If, however, the contribution of neutrons to (\ref{dp}) can be
neglected, then

	\begin{equation}\Sigma_{un}^{(2)}({\bf q})\simeq-{1\over Z}
	{|F_{el}({\bf q})|^2\over G_{E,p}^2(Q_{el}^2)},\label{ds}
	\end{equation}

\noindent completing the analogy of (\ref{dr}) to (\ref{ak}).  For example,
if the contribution of the magnetic ($G_M(Q^2)$) terms is zero, as
discussed by Friar\cite{friar} for a spin-saturated target, then
with $G_{E,n}\!=\!0$, as assumed in (\ref{bn}), (\ref{ds}) is exact.  For
$G_{E,n}\!\not=\!0$, (\ref{ds}) can be suitably modified under reasonable
assumptions, e.g., similar neutron and proton distributions in (\ref{dp}).

The RCSR given here is exact in its treatment of nucleon Fermi motion and
in principle allows the identification of nucleon-nucleon correlations to
arbitrarily high $|{\bf q}|$, given our assumptions about antinucleon and
meson degrees of freedom, nucleon excitations, and off-shell continuation
of nucleon form factors.  From (\ref{dr}), the extraction of the correlation
function $C({\bf q})$ from the calculated Coulomb sum (\ref{dc}) requires the
removal of the nuclear elastic form factor, as just discussed, as well as
a reliable evaluation of the one-body term (\ref{dm}).  For the latter, this
form of the RCSR presupposes that one has in hand the nucleon momentum
distribution $n_\sigma({\bf p})$.

\setcounter{equation}{0}
\seceqe
\section{Expansion in Moments of Nucleon Momentum}

The momentum distribution $n_\sigma({\bf p})$ is not generally available
directly from experimental data.  In order to use the RCSR as given in
the previous section, therefore, one would have to rely on a theoretical
model to calculate the one-body term $\Sigma({\bf q})$.  An alternative
approach, which we develop in this section, is to expand the one-body
term (\ref{dm}) in moments of the nucleon momentum, i.e., in averages of
powers of the nucleon momentum, weighted by the distribution $n_\sigma
({\bf p})$.  The one-body term (\ref{dm}) is then replaced by a sum in terms
of these moments, the first few of which may be known.  This procedure
leads to a series of approximate sum rules, each of which depends on
higher momentum moments. In each case, it is possible to identify appropriate
modifications to the definition of the relativistic Coulomb sum (\ref{dc}) and
arrive at a series of modified sum rules, which share certain features with
the NRCSR.  Of course, such a procedure is useful {\it only\/} if the moment
expansion converges rapidly, and if an adequate number of moments can be
obtained reliably.  In the next section, we will demonstrate that this
expansion indeed converges quickly for a uniform Fermi gas, and argue in
the following discussion that this is likely to be the case in general.

\subsection{Sum Rule I}

The lowest-order momentum moment expansion is obtained by setting
${\bf p}\!=\!0$ in (\ref{dn}):

	\begin{equation}r_\sigma({\bf p},{\bf q})\simeq r_\sigma({\bf 0}
	,{\bf q})={E_{\bf q}+M\over2E_{\bf q}}\ \delta_{\sigma p}
	,\label{ea}\end{equation}

\noindent which is just the factor that appeared in (\ref{ce}) for a single
proton at rest.  This is equivalent to approximating the reduced matrix
element (\ref{df}) by its value at ${\bf p}\!=\!0$:

	\begin{equation}j_{s^\prime s,\sigma}({\bf p},{\bf q})
	\simeq j_{s^\prime s,\sigma}({\bf 0},{\bf q})
	=\sqrt{E_{\bf q}+M\over2E_{\bf q}}\
	\delta_{ss^\prime}\delta_{\sigma p}.\label{eb}
	\end{equation}

\noindent Since we evaluate at ${\bf p}\!=\!0$, (\ref{ea}) and (\ref{eb})
do not involve magnetic terms.  The absence of neutron electric
contributions is a result of our approximation (\ref{bn}).

Inserting (\ref{ea}) into (\ref{dm}), the one-body term becomes

	\begin{equation}\Sigma^{(1)}({\bf q})\simeq Z\ \biggl[
	{E_{\bf q}+M\over2E_{\bf q}}\biggr].\label{ec}\end{equation}

\noindent Similarly, inserting (\ref{eb}) into (\ref{do}) and (\ref{dp}),
the two-body correlation term (\ref{dq}) becomes

	\begin{eqnarray}C({\bf q})&\simeq&
	\biggl[{E_{\bf q}+M\over2E_{\bf q}}\biggr]
	\sum_{{\bf pp}^\prime}\sum_{ss^\prime}\biggl
	\lbrace\langle i|a^\dagger_{{\bf p+q}s}
	a^\dagger_{{\bf p}^\prime s^\prime}
	a_{{\bf p}^\prime+{\bf q}s^\prime}
	a_{{\bf p}s}|i\rangle\nonumber\\
	&&\nonumber\\
	&&\qquad\qquad-\biggl({Z-1\over Z}\biggr)
	\langle i|a^\dagger_{{\bf p}^\prime s^\prime}
	a_{{\bf p}^\prime+{\bf q}s^\prime}|i\rangle
	\langle i|a^\dagger_{{\bf p+q}s}a_{{\bf p}s}
	|i\rangle\biggr\rbrace,
	\label{ed}\end{eqnarray}

\noindent where the proton isospin label has been suppressed for clarity.

Since both $\Sigma^{(1)}({\bf q})$ and $C({\bf q})$ now appear with the
{\it same\/} overall kinematic factor, i.e., that which appeared in (\ref{ce}),
which is a function only of {\bf q}, it is possible to modify the definition
of the Coulomb sum (\ref{dc}) and obtain a more conventional form for the sum
rule.  We define the modified Coulomb sum function:

	\begin{equation}S_I({\bf q})\equiv{\Sigma({\bf q})\over
	r_I({\bf q})}={1\over r_I({\bf q})}\ \int_{\omega_{el}^+}
	^{|{\bf q}|}d\omega\ {W_C(\omega,{\bf q})
	\over G_{E,p}^2(Q^2)},\label{ee}\end{equation}

\noindent where the lowest-order recoil correction factor is

	\begin{equation}r_I({\bf q})\equiv{E_{\bf q}+M\over2E_{\bf q}}
	,\label{ef}\end{equation}

\noindent and $W_C(\omega,{\bf q})$ is to be extracted from experimental
data, using (\ref{aa}).  We then find the {\it approximate\/} sum rule:

	\begin{equation}S_I({\bf q})\simeq Z+{\tilde C}_I({\bf q})
	-{1\over r_I({\bf q})}\biggl[{1\over Z}{|F_{el}({\bf q})|^2
	\over G_{E,p}^2(Q^2_{el})}\biggr].\label{eg}\end{equation}

\noindent This result would be exact for a target of free, stationary
protons.  The two-proton correlation function in (\ref{eg}) is given by

	\begin{eqnarray}{\tilde C}_I({\bf q})&=&
	\sum_{{\bf pp}^\prime}\sum_{ss^\prime}
	\biggl\lbrace\langle i|a^\dagger_{{\bf p+q}s}
	a^\dagger_{{\bf p}^\prime s^\prime}
	a_{{\bf p}^\prime+{\bf q}s^\prime}
	a_{{\bf p}s}|i\rangle\nonumber\\
	&&\nonumber\\
	&&\qquad\qquad-\biggl({Z-1\over Z}
	\biggr)\langle i|a^\dagger_{{\bf p}^\prime s^\prime}
	a_{{\bf p}^\prime+{\bf q}s^\prime}|i\rangle
	\langle i|a^\dagger_{{\bf p+q}s}a_{{\bf p}s}
	|i\rangle\biggr\rbrace,\label{eh}\end{eqnarray}

\noindent in close analogy with (\ref{ai}).  Expression (\ref{eg}) will
be referred to as RCSR-I.

Like the NRCSR given in (\ref{ak}), RCSR-I involves a one-body term $Z$ which
simply counts the total number of charged scatterers, and a two-body term
which contains information on nucleon-nucleon correlations.  The function
${\tilde C}_I({\bf q})$, however, is {\it not\/} the Fourier transform of
the usual spatial correlation function, as defined in terms of local
nucleon field operators, i.e., simply inserting the Dirac nucleon field
operators of (\ref{aab}) into (\ref{ai}) will {\it not\/} yield (\ref{eh}),
because of the antinucleon degrees of freedom which are eliminated from
(\ref{eh}), and the kinematic factors related to relativistic nucleon
recoil which result from that elimination.  However, if the {\it
nonrelativistic\/} result (\ref{ai}) is expressed in momentum space, it
is exactly of the form (\ref{eh}).  Thus this version of the RCSR, although
based on a rather strong assumption about Fermi motion in the target,
is closest in form to the familiar NRCSR.

\subsection{Sum Rule II}

The procedure which led to RCSR-I can be extended to include higher moments
of the nucleon momentum.  In this subsection, we keep terms only through
second order.  We first expand the relativistic recoil factor:

	\begin{equation}r_\sigma({\bf p},{\bf q})=r_0^\sigma+r_1^\sigma\
	({\bf p}\cdot{\bf q})+r_2^\sigma\ {\bf p}^2+r_3^\sigma\
	({\bf p}\cdot{\bf q})^2+{\cal O}({\bf p}^3),\label{ei}\end{equation}

\noindent for isospin projection $\sigma$.  Since the recoil factor
(\ref{dn}) is a function only of {\bf p} and {\bf q}, i.e., not $\omega$,
the expansion coefficients $r_i^\sigma$ are functions only of {\bf q}.
Explicit forms for the $r_i^\sigma$ are given in Appendix B, where we
find $r_0^n\!=\!r_1^n\!=\!0$ as a consequence of approximation (\ref{bn}).

Substituting (\ref{ei}) into the one-body term (\ref{dm}) leads to

	\begin{equation}\Sigma_C^{(1)}({\bf q})\simeq2\sum_{\bf p}
	\biggl\lbrace n_p({\bf p})\Bigl[r_0^p+\Bigl(r_2^p+
	{{\bf q}^2\over3}r_3^p\Bigr){\bf p}^2\Bigr]
	+n_n({\bf p})\Bigl[\Bigl(r_2^n+{{\bf q}^2\over3}
	r_3^n\Bigr){\bf p}^2\Bigr]\biggr\rbrace,\label{ej}\end{equation}

\noindent where we have used the spherical symmetry of $n_\sigma({\bf p})$
to eliminate the $r_1^\sigma$ term and integrate the angles:
$({\bf p}\cdot{\bf q})^2\!\rightarrow\!{1\over3}{\bf p}^2{\bf q}^2$.
We can now identify the average squared nucleon momentum, i.e., the
second momentum moment:

	\begin{equation}\langle{\bf p}^2\rangle_\sigma\equiv
	{1\over N_\sigma}\biggl[2\sum_{\bf p}n_\sigma({\bf p})
	\ {\bf p}^2\biggr],\label{ek}\end{equation}

\noindent where $N_\sigma\!=Z,N$ for protons and neutrons, respectively.
The one-body term (\ref{ej}) can then be expressed

	\begin{equation}\Sigma_C^{(1)}({\bf q})\simeq Z\biggl[r_0^p+
	\Bigl(r_2^p+{{\bf q}^2\over3}r_3^p\Bigr)
	\langle{\bf p}^2\rangle_p\biggr]+N\biggl[
	\Bigl(r_2^n+{{\bf q}^2\over3}r_3^n\Bigr)
	\langle{\bf p}^2\rangle_n\biggr].\label{el}\end{equation}

\noindent We thus have an approximate expression for the one-body term
(\ref{dm}) which depends only on the two parameters $\langle{\bf p}^2
\rangle_\sigma$, and reduces to (\ref{ec}) if $\langle{\bf p}^2\rangle_p
\!=\!\langle{\bf p}^2\rangle_n\!=\!0$.

We can now follow a procedure similar to that of (\ref{ee}), by defining
the modified Coulomb sum function:

	\begin{equation}S_{II}({\bf q})\equiv{\Sigma({\bf q})\over
	r_{II}({\bf q})}={1\over r_{II}({\bf q})}\ \int
	_{\omega_{el}^+}^{|{\bf q}|}d\omega\ {W_C(\omega,{\bf q})
	\over G_{E,p}^2(Q^2)},\label{em}\end{equation}

\noindent where the second-order relativistic recoil factor is

	\begin{equation}r_{II}({\bf q})\equiv\biggl
	[r_0^p+\Bigl(r_2^p+{{\bf q}^2\over3}r_3^p\Bigr)
	\langle{\bf p}^2\rangle_p+{N\over Z}\Bigl(r_2^n+{{\bf q}
	^2\over3}r_3^n\Bigr)\langle{\bf p}^2\rangle_n\biggr].
	\label{en}\end{equation}

\noindent We thus find the second-order {\it approximate\/} sum rule:

	\begin{equation}S_{II}({\bf q})\simeq Z+{\tilde C}_{II}({\bf q})
	-{1\over r_{II}({\bf q})}\biggl[{1\over Z}{|F_{el}({\bf q})|^2
	\over G_{E,p}^2(Q_{el}^2)}\biggr],\label{eo}\end{equation}

\noindent in analogy with (\ref{eg}).  The second-order nucleon-nucleon
correlation function is given by

	\begin{equation}{\tilde C}_{II}({\bf q})={C({\bf q})\over
	r_{II}({\bf q})},\label{ep}\end{equation}

\noindent where $C({\bf q})$ is defined in (\ref{dq}).  Expression
(\ref{eo}) will be referred to as RCSR-II.  Explicit expressions
for the two-body term $\Sigma({\bf q})$ and the correlation function
${\tilde C}({\bf q})$ could be given to the same order of approximation
made in (\ref{ei}), but would require the expansion of $j_{s^\prime s,
\sigma}({\bf p},{\bf q})$ to second-order in {\bf p}.  Since we shall
not use these explicit forms, we do not pursue this here.

Although RCSR-II enjoys a less direct correspondence with the NRCSR than
does RCSR-I, it is more accurate in its treatment of Fermi motion. More
importantly, as we will see in the next section, the higher-order terms
which are included in RCSR-II introduce effects from proton and neutron
anomalous magnetic moments which are appreciable.

\setcounter{equation}{0}
\seceqf
\section{Numerical Results and Test in Fermi Gas Model}

In this Section, we present numerical results for our relativistic Coulomb
sum rules.  We first consider the recoil correction factors $r_I({\bf q})$
and $r_{II}({\bf q})$, which appear in RCSR-I and RCSR-II, respectively,
and find that terms $\sim\langle{\bf p}^2\rangle_\sigma$ in our momentum
expansion bring in anomalous magnetic moment effects which are appreciable.
We then illustrate in the Fermi gas model the accuracy with which RCSR-I
and RCSR-II allow the evaluation of the one-body term.

Ideally, one would like to extract $\langle{\bf p}^2\rangle_\sigma$
directly from experimental data.  Since that is not always possible,
however, one may have to rely upon estimates based on theoretical models.
The simplest such model is a uniform Fermi gas, where $n_\sigma({\bf p})\!
=\!\theta(p_F^\sigma\!-\!|{\bf p}|)$.  In that case, we have:

	\begin{equation}\langle{\bf p}^2\rangle_\sigma
	={3\over5}\ {p_F^\sigma}^2, \label{fa}\end{equation}

\noindent where $p_F^\sigma$ is the Fermi momentum for isospin projection
$\sigma$.  In this example, we assume $Z\!=\!N$ and take $p_F^\sigma\!
=\!p_F\!=\!1.42{\rm\ fm}^{-1}\!=\!.28{\rm\ GeV/c}$ to match the average
density in the interior of finite nuclei.  This gives $\langle{\bf p}^2
\rangle_\sigma\!=\!.047{\rm\ GeV}^2/{\rm c}^2$.  For $Z\!\not=\!N$, it is
reasonable to assume that the Fermi momenta for protons and neutrons scale
according to $p_F^\sigma\!=\!p_F(2N_\sigma/A)^{1/3}$.

Figure 1 shows the recoil correction factors $r_I({\bf q})$ and $r_{II}
({\bf q})$, given in (\ref{ef}) and (\ref{en}), as functions of the
three-momentum transfer {\bf q}.  The dotted curve is $r_I({\bf q})$,
and is indistinguishable (on the scale of this plot) from the factor given
by DeForest: ${\bar G}_{E,p}^2/G_{E,p}^2\!=\!(1\!+\!\tau)/(1\!+\!2\tau)$,
suitably averaged over angle. The dot-dashed curve is $r_{II}({\bf q})$
with $\kappa_p\!=\!\kappa_n\!=\!0$, i.e., for Dirac nucleons.  Both curves
tend to the value $1/2$ as $|{\bf q}|\!\rightarrow\!\infty$.  Their
difference is a measure of the importance of second-order terms in the
moment expansion when anomalous magnetic moments are ignored, and is at
most $\sim\!1/2\%$ anywhere in the range $0\!<\!|{\bf q}|\!<\!\infty$.
The dashed curve is $r_{II}({\bf q})$ with $\kappa_p\!=\!1.79$ and
$\kappa_n\!=\!0$, and shows that the proton anomalous magnetic moment
leads to a noticeable enhancement in the recoil factor: $5\%$ at
$|{\bf q}|\!=\!1{\rm\ GeV}$, and reaching $24\%$ as $|{\bf q}|\!
\rightarrow\!\infty$, when compared to $r_I({\bf q})$.  The solid
curve is $r_{II}({\bf q})$ with $\kappa_p\!=\!1.79$ and $\kappa_n
\!=\!-1.91$, and shows a similar enhancement due to the neutron anomalous
magnetic moment when $N\!=\!Z$: $2\%$ at $|{\bf q}|\!=\!1{\rm\ GeV}$, and
reaching $13\%$ as $|{\bf q}|\!\rightarrow\!\infty$, when compared to
$r_I({\bf q})$.

It is clear from Figure 1 that $r_{II}({\bf q})$ is significantly
different from $r_I({\bf q})$ {\it only\/} when anomalous magnetic
moments are included.  This can be understood by considering the
coefficient of $\langle{\bf p}^2\rangle_p$ in (\ref{en}), i.e., the
quantity $r_2^p\!+\!({\bf q}^2/3)r_3^p$, where $r_2^p$ and $r_3^p$ are
given by (\ref{abc}) and (\ref{abd}), respectively.  We find that the
coefficients of $(1\!+\!\kappa_p)^0$ and $(1\!+\!\kappa_p)^2$ in that
quantity are of roughly the same magnitude and have opposite sign.
Thus the second-order proton contribution to (\ref{en}) is nearly zero
when $\kappa_p\!=\!0$, and $r_{II}({\bf q})\!\simeq\!r_I({\bf q})$.
However, when $\kappa_p\!=\!1.79$, we have $(1\!+\!\kappa_p)^2\!\sim\!8$.
In that case, the coefficients of $(1\!+\!\kappa_p)^0$ and $(1\!+\!
\kappa_p)^2$ no longer cancel, and $r_{II}({\bf q})$ is significantly
enhanced.  Similarly, for neutrons, both $r_2^n$ and $r_3^n$ are positive
and proportional to $\kappa_n^2$.  Thus when $\kappa_n\!=\!-1.91$, we
have $\kappa_n^2\!\sim\!4$ and $r_{II}({\bf q})$ is further enhanced.
{}From the specific functional form of $L_{00}^p({\bf p},q)$, as given in
(\ref{cd}), we note that terms in (\ref{ei}) which are of higher order
in ${\bf p}^2$ will {\it not\/} involve higher powers of the anomalous
terms $(1\!+\!\kappa_p)^2$ and $\kappa_n^2$.  Therefore, we expect that
the moment expansion will converge {\it if\/} the moments themselves
converge.

We now present a numerical test of our moment expansion in the simple system
which includes Fermi motion: a uniform Fermi gas.  The spacelike Coulomb
response function for this system has been studied by several
authors.[13--16, 24]  Substituting (\ref{cc}) into (\ref{ca}),
and integrating over the Fermi sphere $|{\bf p}|\!<\!p_F$ with an appropriate
factor to account for Pauli blocking, we obtain:

	\begin{equation}W_C(\omega,{\bf q})=2\ \sum_\sigma
	\int{d^3p\over(2\pi)^3}n_\sigma({\bf p})
	\bigl[1\!-\!n_\sigma({\bf p}\!+\!{\bf q})\bigr]
	{L_{00}^\sigma({\bf p},{\bf q})\over4E_{{\bf p+q}}E_{\bf p}}
	\delta(\omega-E_{{\bf p+q}}+E_{\bf p}),\label{fb}\end{equation}

\noindent where the occupancy function $n_\sigma({\bf p})\!=\!\theta(p_F\!
-\!|{\bf p}|)$.  Inserting (\ref{fb}) into (\ref{dc}) gives the relativistic
Coulomb sum:

	\begin{equation}\Sigma({\bf q})=2\ \sum_\sigma
	\int{d^3p\over(2\pi)^3}n_\sigma({\bf p})
	\bigl[1\!-\!n_\sigma({\bf p}\!+\!{\bf q})\bigr]
	r_\sigma({\bf p},{\bf q}),\label{fc}\end{equation}

\noindent where we have used definition (\ref{dn}).  Comparison with
(\ref{dm}) makes the identification of one- and two-body contributions
trivial for this model.  Furthermore, the two-body term involves only
Pauli correlations, which vanish for $|{\bf q}|\!>\!2p_F$, and $F_{el}
({\bf q})\!=\!0$ for $|{\bf q}|\!>\!0$.  We now demonstrate in several
cases the accuracy to which $\Sigma({\bf q})$ of (\ref{fc}) is represented
by RCSR-I and RCSR-II.  We take (\ref{fc}) to represent the experimental
Coulomb sum obtained from (\ref{dc}) and (\ref{aa}), and divide out the
recoil factors $r_I({\bf q})$ and $r_{II}({\bf q})$, as in (\ref{ee}) and
(\ref{em}).

Figure 2 shows several versions of the Coulomb sum (divided by $Z$) for
a free Dirac proton gas, i.e., $\kappa_p\!=\!\kappa_n\!=\!0$, versus
three-momentum {\bf q} in units of the Fermi momentum $p_F$.  The dot-dashed
curve is the unmodified Coulomb sum $\Sigma({\bf q})/Z$, given by (\ref{fc});
this is exactly the result given by Matsui,\cite{mat} and tends to the value
$1/2$ as $|{\bf q}|\!\rightarrow\!\infty$.  The dashed curve is $S_I({\bf
q})/Z$, as defined in (\ref{ee}), and the solid curve is $S_{II}({\bf q})/Z$,
as defined in (\ref{em}).  Here, as in Figure 1, the result given by DeForest
is indistinguishable on the scale of this plot from RCSR-I.  Both the dashed
(RCSR-I) and solid (RCSR-II) curves approach unity quickly as $|{\bf q}|\!
\rightarrow\!\infty$, although RCSR-I overshoots slightly ($\sim\!1/10\%$)
in the $1$ GeV region.  The convergence to unity as $|{\bf q}|\!\rightarrow
\!\infty$ is guaranteed in the Dirac case, since both the unmodified Coulomb
sum $\Sigma({\bf q})/Z$ and the recoil factors $r_I({\bf q})$ and $r_{II}
({\bf q})$ approach the value $1/2$, for $\kappa_p\!=\!\kappa_n\!=\!0$.

Figure 3 shows the same quantities for a free proton gas, with $\kappa_p\!
=1.79$.  In this case,  $\Sigma({\bf q})/Z\!\rightarrow\!.62$ as
$|{\bf q}|\!\rightarrow\!\infty$, due to the proton anomalous magnetic
moment.  In contrast to the Dirac case, RCSR-I {\it fails\/} to level off
at $|{\bf q}|\!=\!2p_F$, but rather $S_I({\bf q})/Z\!\rightarrow\!1.23$
as $|{\bf q}|\!\rightarrow\!\infty$.  However, since $r_{II}({\bf q})$
accounts for the anomalous moment, RCSR-II levels off immediately at
$|{\bf q}|\!=\!2p_F$ and $S_{II}({\bf q})/Z\!\rightarrow\!.99$ as
$|{\bf q}|\!\rightarrow\!\infty$.  The difference from unity is a measure
of the importance of higher-order terms, e.g., $\langle{\bf p}^4\rangle
_\sigma$, which have been neglected in our expansion.  Figure 4 shows the same
results for a symmetric ($N\!=\!Z$) Fermi gas.  In this case, neutron terms
further enhance the unmodified sum: $\Sigma({\bf q})/Z\!\rightarrow\!.68$
and $S_I({\bf q})/Z\!\rightarrow\!1.35$ as $|{\bf q}|\!\rightarrow\!
\infty$.  However, when the neutron anomalous moment is incorporated into
the recoil factor $r_{II}({\bf q})$, we again have $S_{II}({\bf q})/Z\!
\rightarrow\!.99$ as $|{\bf q}|\!\rightarrow\!\infty$.

We have shown that it is possible to estimate reliably the one-body term in
the Coulomb sum if one accounts correctly for nucleon recoil.  Due to the
simplicity of the Fermi gas model, however, the numerical study given here
does {\it not\/} test the validity of our nucleons-only approximation, since
the separation of $N$ and ${\bar N}$ degrees of freedom is exact in this case,
nor does it test our assumptions about the off-shell continuation of nucleon
form factors, since $\tau\!=\!{\tilde\tau}$ in this case.  This study does,
however, give an indication of the importance of anomalous magnetic moment
effects, and how accurate these approximate sum rules {\it can\/} be for the
extraction of $NN$ correlation information from electron scattering data,
when the momentum distribution of the nuclear target is like that of a Fermi
gas.

\setcounter{equation}{0}
\seceqg
\section{Discussion and Conclusions}

The main result of this paper is the derivation of a Coulomb sum rule
applicable to the analysis of inelastic electron scattering experiments
on nuclei at high three-momentum transfers.  The restriction to spacelike
four-momenta ($\omega\!<\!|{\bf q}|$), appropriate for electron scattering,
approximately eliminates $N{\bar N}$ pair-production from the response
function.  We then ignore the small contributions which arise due to
antinucleons in the target ground state; this is our ``nucleons-only''
approximation.  The sum rule is given in three forms.

The most general form of the relativistic Coulomb sum rule (RCSR) appears
in (\ref{dr}).  As for the well-known non-relativistic case (NRCSR), the RCSR
can be decomposed into a one-body term ($\Sigma^{(1)}({\bf q})$) and a
two-body correlation term ($C({\bf q})$) after removing the elastic term.
The correlation term is of considerable interest, but is expected to be
dominated by the one-body term for all but the smallest three-momentum
transfers.  In the NRCSR, the one-body term is simply $Z$, the number of
protons.  In the RCSR, however, this term is a function of {\bf q} and
includes a kinematic factor $r_\sigma({\bf p},{\bf q})$, as shown in
(\ref{dm}).  This factor can be interpreted as the relativistic effect of
nucleon recoil, and becomes appreciable for $|{\bf q}|\!\sim\!M$.  In
general, the evaluation of (\ref{dm}) requires knowledge of the full
momentum distribution function $n_\sigma({\bf p})$, which is not accurately
known from experiments.  This then limits the direct applicability of the
RCSR, and in particular, the extraction of correlation information from
the analysis.

We therefore develop an expansion of the one-body term in moments of the
nucleon momentum.  This leads to approximate sum rules, which require only
{\it partial\/} knowledge of the nucleon momentum distribution $n_\sigma
({\bf p})$, i.e., the first few moments, assuming that the moment expansion
is rapidly convergent.  The simplest case is RCSR-I, given in (\ref{eg}),
which involves only the zeroth moment.  This version of the sum rule requires
no information about $n_\sigma({\bf p})$, and therefore has the most direct
correspondence with the familiar NRCSR.  Although RCSR-I is surprisingly
accurate ($\sim\!1/10\%$) in approximating the one-body term for a uniform
system of Dirac protons, it neglects the contributions from the anomalous
magnetic moments, which turn out to be non-negligible ($\sim\!23\%$ for
$|{\bf q}|\!>\!\!>M$ and $Z\!=\!N$).  Therefore, we keep the second moment
contribution to obtain RCSR-II, given in (\ref{eo}), which allows the
evaluation of the one-body term to within $\sim\!1\%$ for a uniform Fermi
gas.  This version of the sum rule requires knowledge of only two parameters,
namely $\langle{\bf p}^2\rangle_\sigma$ for protons and neutrons, which are
approximately known from experimental and theoretical information; we have
used the Fermi gas model for our estimate.  This study shows that the
uncertainty in $\langle{\bf p}^2\rangle_\sigma$ is likely to introduce
much less error into the analysis than would ignoring the anomalous magnetic
moments altogether. The excellent convergence of the moment expansion, shown
here to second-order for a uniform Fermi gas, is not expected to be very
different for correlated systems (see, for example, Donnelly {\it et al.}
\cite{donn}).  Furthermore, the extension of this approach to include
higher moments, if necessary, is straightforward.  We therfore conclude
that RCSR-II, which includes contributions from anomalous magnetic moments,
is the most efficient and reliable method for evaluating the sum rule for
$(e,e^\prime)$ data and separating the two-body information from the
dominant one-body term.

We now discuss the relation of our work to that of earlier authors.
DeForest's\cite{defor} approach to a relativistic sum rule is similar
to that of RCSR-I, both formally and numerically.  The main difference is,
in the language of Section 5, his replacement of $r_\sigma({\bf 0},{\bf q})$
in (\ref{ea}) with the Lorentz invariant factor ${\bar G}_E^2(Q^2)/G_E^2(Q^2)\!
=\!(1\!+\!\tau)/(1\!+\!2\tau)$.  However, a careful inspection of the recoil
factor (\ref{dn}) for a nucleon with arbitrary {\bf p} shows that $r_\sigma
({\bf p},{\bf q})\!\sim\!L_{00}({\bf p},q)/4E_{\bf p+q}E_{\bf p}$ is {\it
not\/} a Lorentz scalar, and that anomalous magnetic moment effects {\it
must\/} enter when ${\bf p}\!\not=\!0$.  Hence there is, in fact, no formal
justification for writing the recoil correction at ${\bf p}\!=\!0$ in
scalar form.  It is interesting that the factor $(1\!+\!\tau)/(1\!+\!2\tau)$
is numerically nearly identical (for all $|{\bf p}|\!\sim\!p_F$) to our
recoil factor $r_I({\bf q})\!\equiv\!r_p({\bf 0},{\bf q})\!=\!(E_{\bf q}\!
+\!M)/2E_{\bf q}$.  However, as we have seen in Figure 1, the correction
due to Fermi motion, which is large due to the effects of anomalous magnetic
moments, must be included for a reliable evaluation of the one-body term.
For example, for $|{\bf q}|\!\sim\!1{\rm\ GeV/c}$, the factor given by
DeForest {\it underestimates\/} relativistic recoil corrections by $\sim\!
7\%$, compared to RCSR-II; therefore, division of the experimental data (as
done in Ref.~{}\cite{mez}, for example) by DeForest's factor in the calculation
of the experimental Coulomb sum will lead to an {\it overenhancement\/} of
the sum by $\sim\!7\%$.

The goal of Donnelly {\it et al.}\cite{donn} is somewhat different: to
find a factor $g(\omega,{\bf q})$ which would replace $G_{E,p}^2(Q^2)$ in
the definition of the spacelike Coulomb sum function, such that the sum will
reach $Z$ in the high-$|{\bf q}|$ limit.  Although their approach can, in
principle, achieve that goal to essentially any order of accuracy, it
necessarily leads to a correction factor $g(\omega,{\bf q})$ which depends
explicitly on the excitation energy $\omega$.  However, unless the
$\omega$-dependence can be extracted from the Coulomb response function
{\it before\/} integration over $\omega$, one will not arrive at a {\it
non-energy-weighted\/} sum rule.  In particular, the approach taken by
Donnelly {\it et al.\/} will necessarily involve energy-weighted sum rules,
which differ formally from the NRCSR, and the RCSR we have derived, in
that dynamical information becomes mixed with the correlation information.

A crucial step in the derivation of our sum rule is the extraction of the
energy dependence from the response function {\it before\/} integration over
$\omega$.  To do this, we needed two assumptions:  first, that the off-shell
form factors could be obtained from the on-shell form factors $G_E(Q^2)$
and $G_M(Q^2)$ as in (\ref{bj}).  Although plausible, this requires theoretical
justification, i.e., a theory of the electromagnetic structure of nucleons.
The second assumption is that the form factors are all proportional, as
in (\ref{bk})--(\ref{bn}).  This is probably sufficiently accurate, except
possibly for taking $G_{E,n}\!=\!0$, as noted earlier. A better approximation
would be to take $G_{E,n}\!\propto\!G_{E,p}$, using data from larger $Q^2$,
although this is incorrect near $Q^2\!=\!0$.

The results of this work do have some bearing on the question of the
saturation of the NRCSR, which we mentioned in Section~1.  The Coulomb
sum function usually extracted from $(e,e^\prime)$ experiments is that
of (\ref{ae}), but with a finite upper limit on $\omega$; since the
theoretical limit at fixed $|{\bf q}|$ is $\omega\!<\!|{\bf q}|$, this
is essentially identical to (\ref{dc}).  For the relativistic sum rules
given in Section 5, one must divide the sum by the relativistic recoil
factor $r({\bf q})$, i.e., either $r_I({\bf q})$ or $r_{II}({\bf q})$,
to obtain a form which should ``saturate'' at $Z$ in the limit that
$|{\bf q}|\!\rightarrow\!\infty$.  Since $r^{-1}({\bf q})\!>\!1$, this
correction will {\it enhance\/} the experimental sum, e.g., as was shown
in Figures 2--4.  At the energies of interest, however, this is only a
small effect, e.g., at $|{\bf q}|\!=\!500{\rm\ MeV/c}$ we have $r_I\!
\simeq\!.94$ and $r_{II}\!\simeq\!.97$.  It is interesting to note that
these factors grow with increasing $\langle{\bf p}^2\rangle$, and
therefore do show {\it in part\/} the increased suppression seen in
larger nuclei.  Of course, many other effects may also contribute to
the observed suppression.

Although we have worked within a ``nucleons-only'' approximation, in
principle, ${\bar N}\!\rightarrow\!{\bar N}$ processes are easily included in
this formalism (since the matrix element of $\Gamma_\mu$ between ${\bar N}$
states is the same as that between $N$ states).  However, we do not pursue
this here because other ${\bar N}$ contributions, e.g., $N{\bar N}$ pair
production, which may also enter the spacelike response, are likely to be
of the same order.  In order to understand fully the contribution of
antinucleons to electron scattering, it will be necessary to study the
spacelike response function in interacting nuclear models, and devise an
appropriate renormalization scheme to obtain finite results.

Two issues which are not treated in this paper must also be addressed
if the relativistic Coulomb sum rule given here is to be used for the
extraction of nucleon-nucleon correlation information: namely, the
contribution of meson exchange currents, and internal excitations of the
nucleon.  The first can be thought of as generating two-body (or more)
terms in the nuclear current operator ${\hat J}_\mu(q)$ in (\ref{bf}), as
done, for example in Ref.~\cite{schi}, where meson current contributions
are actually calculated in a model.  From the point of view of the sum rule,
these many-body contributions can be grouped with the two-body term $\Sigma
^{(2)}({\bf q})$ of (\ref{do}), to be determined {\it experimentally\/} as
what is left over when $\Sigma^{(1)}({\bf q})$ is subtracted from the
Coulomb sum.  The second issue can be thought of, to a first approximation,
as the problem of removing the ``background'' of $N(e,e^\prime)N^*$
excitations from the $(e,e^\prime)$ response, to leave the ``nucleons-only''
response, to which our relativistic Coulomb sum rules apply.  This could
be done theoretically by calculating the $N^*$ background, e.g., in a
noninteracting-nucleus model.  Alternatively, some partially exclusive
experimental information could help: for example, $(e,e^\prime p)$ data
in the quasi-free peak region can be used to normalize the knockout
of protons (as opposed to $N^*$), although other process (e.g., $\Delta
\!\rightarrow\!p\!+\!\pi$) will also contribute to this process.  These
two problems, and the problem of antinucleons, are subjects for future
investigation.

\vskip 0.2 true in
{\bf Acknowledgements}
\vskip 0.2 true in

This research was supported in part by the U.~S.~Department of Energy
under Grant No.~DE-FG02-88ER40425 with the University of Rochester.
We would also like to thank the High Energy Physics Group for the
use of their VAX computer.

\setcounter{equation}{0}
\seceqaa
\appendix
\section{Single-fermion Solutions}

The Dirac equation for free point nucleons represented by the wavefunction
$\psi(x)$ can be written

	\begin{equation}\Bigl(i\gamma^\mu\partial_\mu-M\Bigr)
	\psi(x)=0,\label{aaa}\end{equation}

\noindent where $x^\mu\!=\!(t,{\bf x})$.  For a static, spatially uniform
system, the solutions to (\ref{aaa}) can be expressed as momentum eigenstates
with momentum {\bf p}.  The energy eigenvalues for positive and negative
energy solutions are $\epsilon^{(\pm)}_{\bf p}\!=\!\pm E_{\bf p}$, where
$E_{\bf p}\!\equiv\!\sqrt{{\bf p}^2+M^2}$.

{}From the complete set of solutions to (\ref{aaa}), it is possible to
construct
a local field operator in terms of nucleon and antinucleon degrees of
freedom:

	\begin{equation}{\hat\psi}({\bf x})=
	{1\over\sqrt{V}}\sum_{{\bf p}s}
	\biggl[{u_s({\bf p})\over\sqrt{2E_{\bf p}}}
	\ {\rm e}^{i{\bf p\cdot x}}\ a_{{\bf p}s}
	+{v_s({\bf p})\over\sqrt{2E_{\bf p}}}
	\ {\rm e}^{-i{\bf p\cdot x}}\ b^\dagger_{{\bf p}s}
	\biggr],\label{aab}\end{equation}

\noindent where $V$ represents the normalization volume.  The nucleon
and antinucleon spinors are defined

	\begin{equation}u_s({\bf p})=\sqrt{E_{\bf p}+M}
	\Biggl[{\chi_s\atop{{\bf \sigma\cdot p}\over
	E_{\bf p}+M}\chi_s}\Biggr]\qquad
	v_s({\bf p})=\sqrt{E_{\bf p}+M}
	\Biggl[{{{\bf \sigma\cdot p}\over
	E_{\bf p}+M}\chi_s\atop\chi_s}\Biggr]
	\label{aac}\end{equation}

\noindent respectively, and have been normalized to $2E_{\bf p}$ particles
per unit volume.  Here $\chi_s$ is the usual two-dimensional Pauli spinor,
and $a_{{\bf p}s}$ and $b^\dagger_{{\bf p}s}$ are destruction and
creation operators for nucleon and antinucleons, respectively.

It is possible to generalize this formalism to include isospin
degrees of freedom.  We simply define $u_{s\sigma}({\bf p})\!\equiv
\!u_s({\bf p})\eta_\sigma$, where $\eta_\sigma$ is a two dimensional
spinor corresponding to isospin projection $\sigma$, and include a sum
over $\sigma$ in (\ref{aab}).

\setcounter{equation}{0}
\seceqab
\section{Coefficients of Moment Expansion}

The expansion coefficients appearing in (\ref{ei}) are:

	\begin{equation}r_0^p={E_{\bf q}+M\over2E_{\bf q}}
	\label{aba}\end{equation}

	\begin{equation}r_1^p={{\bf q}^2\over2ME_{\bf q}^3}
	\label{abb}\end{equation}

	\begin{equation}r_2^p={-{\bf q}^4(2E_{\bf q}+M)\over
	4M^2E_{\bf q}^3(E_{\bf q}+M)^2}
	+{(1+\kappa_p)^2{\bf q}^2\over
	2M^2E_{\bf q}(E_{\bf q}+M)}\label{abc}\end{equation}

	\begin{equation}r_3^p={2E_{\bf q}^4-M(E_{\bf q}+M)
	(2E_{\bf q}^2-3M^2)\over4M^2E_{\bf q}^5(E_{\bf q}+M)}
	-{(1+\kappa_p)^2\over2M^2E_{\bf q}(E_{\bf q}+M)}
	\label{abd}\end{equation}

	\begin{equation}r_2^n={\kappa_n^2{\bf q}^2\over2M^2E_{\bf q}
	(E_{\bf q}+M)}\label{abe}\end{equation}

	\begin{equation}r_3^n={-\kappa_n^2\over2M^2E_{\bf q}
	(E_{\bf q}+M)}\label{abf}\end{equation}

\noindent We have $r_0^n\!=\!r_1^n\!=\!0$ by our approximation
$G_{E,n}(Q^2)\!=\!0$.

Expressions (\ref{abe}) and (\ref{abf}) can be obtained from (\ref{abc}) and
(\ref{abd}), respectively, by keeping only terms proportional to $(1\!+\!
\kappa_p)^2$ and letting $(1\!+\!\kappa_p)^2\!\rightarrow\!\kappa_n^2$.  The
corresponding expressions for Dirac nucleons are obtained by setting
$\kappa_p\!=\!\kappa_n\!=\!0$.

\vfill
\eject

\begin{figure}
\caption{Recoil correction factors vs. three-momentum transfer
${\bf q}$.  The curves representing $r_I({\bf q})$ and $r_{II}({\bf q})$
with $\kappa_p\!=\!\kappa_n\!=\!0$ are nearly indistinguishable on this
scale.}
\label{fig1}
\end{figure}

\begin{figure}
\caption{Relativistic Coulomb sum (divided by $Z$) vs. three-momentum
transfer {\bf q} in the Fermi gas model, ignoring anomalous magnetic
moments.  The dashed curve is $S_I({\bf q})/Z$, and the solid curve
is $S_{II}({\bf q})/Z$.  For comparison, the dot-dashed curve is the
unmodified sum $\Sigma({\bf q})$/Z.}
\label{fig2}
\end{figure}

\begin{figure}
\caption{Relativistic Coulomb sum (divided by $Z$) vs. three-momentum
transfer ${\bf q}$ in the Fermi gas model, including the proton anomalous
magnetic moment.  The dashed curve is $S_I({\bf q})/Z$, and the solid
curve is $S_{II}({\bf q})/Z$.  For comparison, the dot-dashed curve is
the unmodified sum $\Sigma({\bf q})$/Z.}
\label{fig3}
\end{figure}

\begin{figure}
\caption{Relativistic Coulomb sum (divided by $Z$) vs. three-momentum
transfer ${\bf q}$ in the Fermi gas model, including proton and neutron
anomalous magnetic moments.  The dashed curve is $S_I({\bf q})/Z$, and
the solid curve is $S_{II}({\bf q})/Z$.  For comparison, the dot-dashed
curve is the unmodified sum $\Sigma({\bf q})$/Z.}
\label{fig4}
\end{figure}

\vfill
\eject


\begin{thebibliography}{99}
\bibitem{mvvh}K.~W.~McVoy and L.~Van~Hove, Phys.~Rev.~{\bf 125}, 1034 (1962).
\bibitem{xa}Z.-E.~Meziani {\it et al.}, Nucl.~Phys.~{\bf A446}, 113 (1985).
\bibitem{xb}R.~Altemus {\it et al.}, Phys.~Rev.~Lett.~{\bf 44}, 965 (1980).
\bibitem{xc}P.~Barreau {\it et al.}, Nucl.~Phys.~{\bf A358}, 287 (1981).
\bibitem{xd}P.~Barreau {\it et al.}, Nucl.~Phys.~{\bf A402}, 515 (1983).
\bibitem{xe}M.~Deady {\it et al.}, Phys.~Rev.~C~{\bf 28}, 631 (1983).
\bibitem{xf}Z.-E.~Meziani {\it et al.}, Phys.~Rev.~Lett.~{\bf 52}, 2130
	(1984).
\bibitem{xg}M.~Deady {\it et al.}, Phys.~Rev.~C~{\bf 33}, 1897 (1986).
\bibitem{pand}V.~R.~Pandharipande, Nucl.~Phys.~{\bf A497}, 43c (1989).
\bibitem{pandb}O.~Benhar, V.~R.~Pandharipande and S.~C.~Pieper,
	Rev.~Mod.~Phys.~{\bf 65}, 817 (1993).
\bibitem{schi}R.~Schiavilla, R.~B.~Wiringa and J.~Carlson,
	Phys.~Rev.~Lett.,{\bf 70}, 3856 (1993).
\bibitem{wal}J.~D.~Walecka, Nucl.~Phys.~{\bf A399}, 387 (1983).
\bibitem{mat}T.~Matsui, Phys.~Lett.~B~{\bf 132}, 260 (1983).
\bibitem{ddb}G.~Do~Dang, M.~L'Huillier, Nguyen~Van~Giai and
	J.~W.~Van~Orden, Phys.~Rev.~C~{\bf 35}, 1637 (1987).
\bibitem{hor}C.~J.~Horowitz, Phys.~Lett.~B~{\bf 208}, 8 (1988).
\bibitem{donn}T.~W.~Donnelly, E.~L.~Kronenberg and J.~W.~Van~Orden,
	Nucl.~Phys.~{\bf A494}, 365 (1989).
\bibitem{defor}T.~De~Forest,~Jr., Nucl.~Phys.~{\bf A414}, 347 (1984).
\bibitem{mez}Z.-E.~Meziani {\it et al.}, Phys.~Rev.~Lett.~{\bf 69}, 41
	(1992).
\bibitem{pb}M.~A.~Preston and R.~K.~Bhaduri, {\it The Structure of the
	Nucleus\/} (Addison-Wesley, Reading, Mass., 1975).
\bibitem{hoh}G.~Hohler {\it et al.}, Nucl.~Phys.~{\bf B114}, 505 (1976).
\bibitem{vac}R.~J.~Perry, Phys.~Lett.~B~{\bf 182}, 269 (1986).
\bibitem{spf}R.~Schiavilla, V.~R.~Pandharipande and A.~Fabrocini,
	Phys.~Rev.~C~{\bf 40}, 1484 (1989).
\bibitem{friar}J.~L.~Friar, Ann.~Phys.~{\bf 81}, 332 (1973).
\bibitem{dda}G.~Do~Dang and Pham~Van~Thieu, Phys.~Rev.~C~{\bf 28},
	1845 (1983).
\end{thebibliography}
\end{document}